\documentclass[12pt]{article}
\usepackage{bbm}
\usepackage{t1enc}
\usepackage[latin2]{inputenc}
\usepackage{amsmath,amssymb,amsthm}
\usepackage{txfonts}
\usepackage{sectsty}
\usepackage{overcite}
\usepackage{mathrsfs}
\usepackage{fixmath}
\usepackage[margin=1in]{geometry}
\usepackage{titlesec}
\usepackage{upgreek}

\newcommand{\eq}[1]{\begin{equation}#1\end{equation}}
\newcommand{\define}{\coloneqq}
\newcommand{\cd}{\!\cdot\!}
\newcommand{\hh}{\mathscr{H}}
\newcommand{\nn}{\mathrm{N}}

\newcommand{\g}{\boldsymbol{g}}
\newcommand{\pp}{\boldsymbol{p}}
\newcommand{\ppi}{\boldsymbol{\pi}}
\newcommand{\G}{\Upgamma}
\newcommand{\cc}[1]{{C^\infty}_{\!\!\!\!\!{\scriptscriptstyle 0}\,\,}(#1)}
\newcommand{\ccc}{{C^\infty}_{\!\!\!\!\!{\scriptscriptstyle 0}\,\,}}
\newcommand{\sob}[1]{{H^{#1}}_{\!\!\!{\scriptscriptstyle 0\,}}}
\newcommand{\scc}[1]{{C^{\infty}}_{\!\!\!\!\!\!{\scriptscriptstyle\mathrm{o}}\,\,}(#1)}

\newcommand{\ah}{\alpha^\mu\cd\hh_\mu}
\newcommand{\nh}{\nn^\mu\cd\hh_\mu}
\newcommand{\f}{\boldsymbol{\phi}}
\newcommand{\p}{\boldsymbol{\psi}}
\newcommand{\supp}{\mathop{\mathtt{supp}}}
\newcommand{\dom}{\mathop{\mathrm{dom}}}
\newcommand{\ran}{\mathop{\mathrm{ran}}}

\newcommand{\midint}{\mathop{\mbox{$\int$}}}
\newcommand{\midsum}{\mathop{\mbox{$\sum$}}}
\newcommand{\cO}{\bar{\Omega}}
\newcommand{\D}{\mathbold{D}}
\newcommand{\cD}{\bar{\D}}
\newcommand{\E}{\mathbold{E}}
\newcommand{\cE}{\bar{\E}}
\newcommand{\K}{\mathbold{K}}
\newcommand{\cK}{\bar{\K}}
\newcommand{\sS}{\mathbold{S}}
\renewcommand{\j}{\!j}
\renewcommand{\d}{\mathrm{d}}    
\renewcommand{\Omega}{\Upomega}
\renewcommand{\Sigma}{\Upsigma}
\renewcommand{\S}{\Sigma}
\renewcommand{\O}{\Omega}
\renewcommand{\leq}{\leqslant}
\renewcommand{\le}{\leqslant}

\renewcommand{\geq}{\geqslant}

\theoremstyle{plain}\newtheorem{theor}{Theorem}
\theoremstyle{plain}\newtheorem*{prop}{Proposition} 
\theoremstyle{plain}\newtheorem{lem}{Lemma}
\theoremstyle{remark}
\theoremstyle{plain}\newtheorem*{cor}{Corollary}
\theoremstyle{definition}\newtheorem*{definition}{Definition}
\theoremstyle{plain}\newtheorem*{theorname}{Theorem}

\newcommand{\refonline}[1]{[\citeonline{#1}]}
\newcommand{\affiliation}[1]{
              \noindent\hspace*{2.5em}\parbox[t]{30em}{\it#1}
              \vspace*{2ex}\\}
\titlelabel{\thetitle.\hspace{8pt}}
\renewcommand\thesection{\Roman{section}}

\renewcommand{\title}[1]{{\Large\bf\flushleft{#1}}\vspace*{3ex}\\}
\renewcommand{\author}[2]{{\noindent\hspace*{2.5em}\large#1}
                     \footnote{Electronic mail: $\mathtt{#2}$}\\}
\sectionfont{\normalsize\bf}
\subsectionfont{\normalsize\bf}
\makeatletter\renewcommand\@biblabel[1]{$^{#1}$}\makeatother
\begin{document}
\bibliographystyle{utphys}
\title{Gravity from the extension of spatial diffeomorphisms}

\author{Szilard Farkas}{farkas@uchicago.edu}
\author{Emil J. Martinec}{ejmartin@uchicago.edu}

\affiliation{Enrico Fermi Institute and Department of Physics, University of Chicago,\\ Chicago, IL 60637, USA}
\vspace{-5ex}

\begin{abstract}
The possibility of the extension of spatial diffeomorphisms to a larger family of symmetries in a class of classical field theories is studied. The generator of the additional local symmetry contains a quadratic kinetic term and a potential term which can be a general (not necessarily local) functional of the metric. From the perspective of the foundation of Einstein's gravity our results are positive: The extended constraint algebra is either that of Einstein's gravity, or ultralocal gravity. If our goal is a simple modification of Einstein's gravity that for example makes it perturbatively renormalizable, as has recently been suggested, then our results show that there is no such theory within this class.
\end{abstract}
\section{Introduction}
What makes an $n+1$ dimensional (pseudo) Riemannian manifold the relevant structure in the description of the time evolution of Riemannian $n$-geometries? In the Hamiltonian formalism, which is based on a spacelike foliation of the spacetime, the invariance of Einstein's gravity under nonspatial diffeomorphisms -- diffeomorphisms that cannot be restricted to the spatial slices of the foliation -- is somewhat hidden, whereas the spatial diffeomorphism symmetry is as manifest as in the Lagrangian formalism. More concretely, it requires some calculation to realize that the infinitesimal symmetries that the Hamiltonian constraint generates on the solutions to the equations of motion correspond to infinitesimal deformations of the foliation. Since the relationship between the transformations generated by the constraints and the diffeomorphisms of the underlying $n+1$ dimensional manifold is not straightforward, the question naturally arises if this relationship is necessary at all. To put it differently: Do spatial diffeomorphisms allow for a further local symmetry so that the symmetry algebra is different from that of general relativity or ultralocal gravity? We insist on general covariance, however, we use this term in a weaker sense than usual: The spatial metric and the conjugate momentum are the only canonical variables, and the local constraints are constructed out of them in a form invariant way.

This is not the first time that the possibility of reducing the set of postulates that lead to Einstein's gravity is considered.\cite{teitelboim,HKT,kuchar,BFM} However, we are interested in this question not only from the perspective of the foundation of Einstein's gravity. Another motivation is a recent proposal by Ho\v{r}ava for a modified theory of gravity in which Lorentz invariance is given up in the quest for an improved UV behavior.\cite{horava} Although this model does not seem to be physically relevant, it puts our question in a new light: Perhaps it is possible to reduce the postulates of general relativity so that they are realized not only by general relativity and ultralocal gravity, but also by some other theories which might have the properties that motivated Ho\v{r}ava's modification. First we briefly describe Ho\v{r}ava's proposal, and set up the framework in which we can look for theories that can be relevant for gravity.        

The starting point of Ho\v{r}ava's construction is the action of general relativity in terms of the ADM Lagrangian:
\eq{
\label{adm_action}
S_\mathrm{ADM}=\midint_{\mathbb{R}\times\Sigma}\mathscr{L}_{\mathrm{ADM}}=\midint_{\mathbb{R}\times\Sigma}\nn\sqrt{g}(R+K_{i\j}K^{i\j}-K^2),\quad\quad K_{i\j}=\frac{1}{2\nn}(\dot{g}_{i\j}-2\nabla_{(i}\nn_{\j)}),
}
where $\Sigma$ is the space, and $\mathbb{R}$ is the time. The field configuration on the spacetime manifold is given by $(g_{i\j},\nn^\mu)$, $\nn^\mu=(\nn,\nn^i)$, where $\nn$ is the lapse, $\nn^i$ are the shift functions, and $g_{i\j}$ is the spatial metric. In \eqref{adm_action} $R$ is the Ricci scalar of $g_{i\j}$, and $\nabla$ is the covariant derivative on $\Sigma$ compatible with the spatial metric. $K_{i\j}$ and $K$ are the extrinsic curvature and its trace, and $\dot{g}_{i\j}$ is the time derivative of $g_{i\j}$. Latin indices denote spatial components. A configuration in the phase space $\G$ is given by $(g_{i\j},K_{i\j})$. To avoid indices, we will denote the elements of $\G$ by $(\g,\ppi)$, where $\ppi$ is the momentum conjugate to $\g$. 

Spatial diffeomorphisms are still local symmetries in Ho\v{r}ava's proposal as in Einstein's gravity. What remains from the nonspatial diffeomorphisms is the invariance of the action under space independent, i.e.\ foliation preserving transformations. For further restrictions on the possible action, Ho\v{r}ava first appeals to an effective field theory argument: Under an anisotropic scaling symmetry, the theory has a UV fixed point, and apart from the UV potential the Lagrangian contains only relevant couplings under the scaling symmetry. For pragmatic reasons, a further principle is introduced. The theory should satisfy a ``detailed balance'' condition, which restricts the form of the potential term in the action: 
\eq{
\label{hr_action}
S=\int\d t\,\d^3\!x\,\nn\sqrt{g}\left(\frac{1}{\kappa^2}K_{i\j}G^{i\j kl}K_{kl}-\frac{\kappa^2}{16}\frac{\delta W}{\delta g_{i\j}}G_{i\j kl}\frac{\delta W}{\delta g_{kl}}\right),
}
where $W$ and the tensor $G^{i\j kl}$ are some functionals of the spatial metric, and $\kappa$ is a constant. The UV potential is proportional to $C^{i\j}C_{i\j}$, where $C^{i\j}$ is the Cotton tensor. The detailed balance condition limits the number of the possible relevant couplings under the anisotropic scaling symmetry if we assume that this condition is preserved by the renormalization group flow. The action suggested by Ho\v{r}ava is \eqref{hr_action} with 
\begin{align}
\label{gen_wheeler}
&G^{i\j kl}=g^{ik}g^{jl}+g^{il}g^{jk}-\lambda\,g^{i\j}g^{kl},\\&W=\frac{1}{w^2}\int\d^3\!x\,\mathop{\mathrm{Tr}}\bigg(\Gamma\wedge d\Gamma+\frac{2}{3}\Gamma\wedge\Gamma\wedge\Gamma\bigg)+\mu\int\d^3\!x\,\sqrt{g}(R-2\Lambda_W),\label{W}
\end{align} 
where $\lambda$, $w$, $\mu$, and $\Lambda_W$ are constants, and the first term in $W$ is the gravitational Chern-Simons term, expressed in terms of the Christoffel symbols ${\Gamma^i}_{\!\!\j k}$ of $g_{i\j}$. The functional derivative of this term with respect to the metric is proportional to $C^{i\j}$. 

By discarding possible spatial boundary terms, \eqref{hr_action} and its later generalizations can be got from the following action given in the Hamiltonian formalism:
\eq{
\label{action_ham_gen}
S=\midint_{\;t_1}^{\;\;t_2}\d t\midint_\S\Big(\dot{\g\phantom{.}}\!\cd\ppi-\nn^\mu\hh_\mu[\g,\ppi]\Big), 
}
where at any time $\hh_\mu=(\hh,\hh_i)$ are functionals of the spatial configurations of $(\g,\ppi)$. This action is defined on a class of time dependent configurations $(t,x)\mapsto\big(\g(t,x),\ppi(t,x),\nn^\mu(t,x)\big)$, $t\in(t_1,t_2)$, $x\in\S$. Any field that can be an instantaneous configuration in a function within this class will be referred to as kinematically possible. A field that can occur as an instantaneous configuration in a (local) solution to the equations of motion of \eqref{action_ham_gen} will be called dynamically possible. Let $\hat{\G}$ be the constraint submanifold of $\G$, which, by definition, consists of the dynamically possible canonical variables. We will assume that the infinitesimal transformations generated by first class constraints on $\hat{\G}$ can be integrated to a symmetry group $\mathrm{CAN}$. Subscript $\mathrm{ADM}$ will denote the corresponding objects in general relativity.

In the nonprojectable version of Ho\v{r}ava's theory $\nn$ can be any smooth function apart from some conditions on its asymptotic behavior. In this version the equations of motion include the constraints $\hh_\mu=0$ on $\G$, which are not first class.\cite{li_pang} We mention three different approaches to this problem: 
\vspace{0.5ex}

\noindent(i) We can get first class constraints by imposing further constraints on $\G$. This can be implemented by adding constraint functions with multipliers to \eqref{action_ham_gen}. The new symmetry group $\mathrm{CAN}'$ is an extension of the spatial diffeomorphisms by local symmetries. The new constraint submanifold $\hat{\G}'\subsetneqq\hat{\G}$ is a submanifold of $\hat{\G}$, so $\hat{\G}'$ and especially the manifold $\hat{\G}'/\mathrm{CAN}'$ of orbits are too small compared to these objects in general relativity. Thus this option cannot be relevant for gravity.

\vspace{0.5ex}
\noindent(ii) The nonspatial diffe\-o\-mor\-phisms in Einstein's theory are symmetries of \eqref{hr_action} only if they preserve the foliation. These symmetries can be implemented on the action if the set of kinematically possible configurations is invariant under such transformations. For example $\nn$ can be kinematically possible only if it is a constant function, and thus the spacetime configurations of $\nn$ depend only on time. This is also called the projectable version of the theory. Here the constraints are first class since apart from the constraints implied by the spatial diffeomorphism symmetry, there is one single global constraint which is invariant under the transformations of the canonical variables by spatial diffeomorphisms.

\vspace{0.5ex}
\noindent(iii) It is not necessarily a pathology if the constraints are not first class. It just means that $(\g,\ppi,\nn^\mu)$ is not dynamically possible for any kinematically possible $\nn$ and $(\g,\ppi)\in\hat{\G}$. It is possible that we do not have to modify the theory as in (i) if our only goal is that the local constraints $\hh_\mu=0$ are preserved by a nontrivial time evolution, for which all the constraints need to be satisfied by a nonzero $\nn$ for any $(\g,\ppi)\in\hat{\G}$. A dynamically possible configuration of $\nn[\g_0,\ppi_0]$ for a given $(\g_0,\ppi_0)\in\hat{\G}$ satisfies $\{\,\hh[\g,\ppi](x)\,,\smallint_\S\nn[\g_0,\ppi_0]\hh[\g,\ppi]\,\}|_{\g=\g_0,\ppi=\ppi_0}=0$, where the Poisson bracket is taken with respect to $(\g,\ppi)$. On the other hand, it is necessary that $\nn[\g_0,\ppi_0]$ satisfies this equation in order for $\delta F[\g_0,\ppi_0]=\{\,F[\g,\ppi]\,,\smallint_\S\nn[\g_0,\ppi_0]\hh[\g,\ppi],\}|_{\g=\g_0,\ppi=\ppi_0}$ to define  an infinitesimal symmetry on $\hat{\G}$. It was argued in \refonline{henneaux} that in a class of theories which includes Ho\v{r}ava's proposal the only solution to this equation is $\nn[\g_0,\ppi_0]=0$ for generic coupling constants and $(\g_0,\ppi_0)\in\hat{\G}$. Thus there is no symmetry associated with $\hh$.

\vspace{0.5ex}

There are several reasons why the lack of the local constraint $\hh=0$ or the corresponding local symmetry is undesirable. If $\lambda\neq\frac{2}{3}$ in \eqref{gen_wheeler} then the local scale factor of the spatial metric is dynamical, i.e., using the variables $\phi\define\frac{1}{3}\ln g$ and $\tilde{g}_{i\j}\define g_{i\j}e^{-\phi}$ instead of $g_{i\j}$ in \eqref{hr_action}, we can see that $K_{i\j}G^{i\j kl}K_{kl}$ is independent of $\phi$ if $\lambda=\frac{2}{3}$, whereas it is quadratic in $\dot{\phi}$ for any other $\lambda$, and the acceleration of $\phi$ appears in the field equations. If $\lambda>\frac{2}{3}$ in \eqref{gen_wheeler}, then the signature of $G^{i\j kl}$ as a metric on the symmetric rank two tensors is $(-,+,+,+,+,+)$, and the term in \eqref{hr_action} proportional to $\dot{\phi}^2$ is negative. Such a mode usually leads to the loss of unitarity in quantum theory unless it is unphysical, i.e., there are gauge symmetries, and there is such a gauge condition that fixes this mode apart from some nonlocal degrees of freedom, which is not the case if the only local symmetries are the spatial diffeomorphisms. In Einstein's gravity it is the space dependent nonspatial diffeomorphisms, which are generated by the Hamiltonian constraint, that make the gauge choice ${K^i}_{\!\!i}=0$ admissible at least for spatial metrics that satisfy some asymptotic conditions in suitably chosen coordinates (Dirac's maximal slicing gauge, see for example \refonline{HRT}). Note that $\lambda$ changes between $\lambda=\frac{2}{3}$ and $\lambda=1$ if general relativity is to be recovered from \eqref{hr_action} in some limit, and the local scale factor of the metric is a physical mode with negative kinetic energy for such $\lambda$ if there is no symmetry that replaces the temporal diffeomorphisms. Even if it was possible to define a unitary quantum theory with such a physical mode, its relationship with Einstein's gravity, where this mode is unphysical, would be unclear.

The lack of a local symmetry that replaces space dependent temporal diffeomorphisms also makes the physical interpretation of certain solutions problematic. Let $\mu=0$ in the potential \eqref{W}, and let us break the detailed balance condition, which seems to be necessary in order that general relativity be some limit of the modified theory.\cite{nastase} If we add the Ricci scalar of the spatial metric to \eqref{hr_action} so that the limit $w\to\infty$ yields the ADM action \eqref{adm_action}, we get
\eq{
\label{hr_action2}
S=\int\d t\,\d^3\!x\,\nn\sqrt{g}\left(\frac{1}{\kappa^2}K_{i\j}G^{i\j kl}K_{kl}-\frac{\kappa^2}{2w^4}C^{i\j}C_{i\j}+\frac{\kappa^2}{2}R\right),
}
where we used that $C^{i\j}$ is traceless, so $G_{i\j kl}C^{i\j}C^{kl}=2C^{i\j}C_{i\j}$. A spherically symmetric spatial metric is conformally flat, and the Cotton tensor is zero for such metrics in three dimensions. For $(g_{i\j},\nn,0)$, where $g_{i\j}$ is conformally flat and $\dot{g}_{i\j}=0$, the field equations of \eqref{hr_action2} reduce to the same equations as those of the ADM action \eqref{adm_action}. So the configuration $(g_{i\j},\nn,\nn^i)$ obtained from the Schwarzschild spacetime metric in the Schwarzschild coordinates is a solution to the field equations of \eqref{hr_action2}. The horizon is a coordinate singularity in Einstein's gravity, but here it appears to be a physical singularity, where the spatial metric changes its signature. In general relativity the coordinate singularities can be removed locally by taking a family of timelike geodesics heading towards the singularity, and using their proper time as the new time coordinate. Such a family of timelike geodesics can be arbitrarily approximated by physically realizable observers (by some matter distribution). In Ho\v{r}ava's theory, with spatial diffeomorphisms as the only local symmetries, there is a preferred foliation of the spacetime. The label of the spacelike hypersurfaces serves as a time coordinate. If this time coordinate has any physical meaning in the sense that there are physically realizable observers whose time arbitrarily can approximate it, the spatial metric changes its signature at some points according to such observers. This phenomenon is hard to interpret. 

Of course it is possible that such observers are not realizable. For example the acceleration of a stationary particle may be unbounded as its radial coordinate approaches the location of the horizon. Note that even if this is true in Einstein's gravity, it does not necessarily hold in Ho\v{r}ava's modification. It is not obvious how Ho\v{r}ava's proposal incorporates matter, and if test particles respect the geodesic principle. The geodesic principle is true if the tensor whose nonzero value indicates the presence of matter is obtained by varying a diffeomorphism invariant action with respect to the metric, and it satisfies the dominant energy condition. (See \refonline{ehlers_geroch} for the details, including the precise meaning of the geodesic principle here.) Even if we reconstruct a spacetime metric from the field variables in Ho\v{r}ava's theory, this result does not apply. Nevertheless, assume that the geodesic principle holds. It is a natural requirement that there be a time coordinate which can be measured by a procedure which is physically admissible in the entire space that should contain the region where the spatial metric is properly Riemannian. So we are in the puzzling situation that the acceleration of stationary test particles can be arbitrarily high in any time coordinate that is defined by physically feasible instructions. 

As mentioned earlier, Ho\v{r}ava proposed a UV theory that has a local scale invariance. The action is \eqref{hr_action} with only the least relevant potential term:
\eq{
\label{UV}
S_{\mathrm{UV}}=\int\d t\,\d^3\!x\,\nn\sqrt{g}\left\{\frac{2}{\kappa^2}\left(K_{i\j}K^{i\j}-\frac{1}{3}K^2\right)-\frac{\kappa^2}{2w^4}C^{i\j}C_{i\j}\right\},
}
which is invariant under $(g_{i\j},\nn,\nn^i)\to(g_{i\j}e^{2\omega},\nn e^{3\omega},\nn^i)$, where $\omega$ is an arbitrary smooth function on $\Sigma$. The Cotton tensor transforms as $C^{i\j}\to C^{i\j}e^{-5\omega}$. In order to implement this symmetry, the space of the kinematically possible configurations of $\nn$ should be invariant under the scale transformation. This space is big enough to get a local constraint by varying the action with respect to $\nn$. The constraints on the phase space are not first class (see \refonline{li_pang} for direct computations for the special case when the action is the same as \eqref{UV} except for the kinetic term, which is taken to be the same as in the ADM action \eqref{adm_action}). It is a relevant question what generalizations of the initial class of theories might allow for a first class constraints, or if it is possible to realize the local scale transformation so that this additional local constraint is absent.

We have argued that the plausible candidates for a modification of Einstein's gravity seem to be the ones that admit an extension of the spatial diffeomorphisms with a further local symmetry. Actually, this could be the guiding principle in our construction of theories in which only the spatial diffeomorphisms are left from the symmetries of general relativity: We require that the temporal diffeomorphisms are replaced with a new local symmetry. We have shown above some physical motivations for this requirement, and we think it should replace the detailed balance condition whose status in Ho\v{r}ava's original proposal is dubious. It was introduced only to avoid the proliferation of the possible terms in the Lagrangian, and in fact it turns out to be too restrictive if the theories that respect it are designed to reproduce Einstein's gravity in some limit. Breaking the detailed balance condition in order to overcome the latter difficulty raises the question that the imposition of this condition initially intended to answer: Why do we not have to switch on the other relevant terms that break the detailed balance condition? There are intimidatingly many such terms. 

Once we accept the need for extensions of spatial diffeomorphisms, we have to choose a framework in which they are easy to find. The symmetries of Einstein's gravity take a simple form in the Lagrangian formalism. They are diffeomorphisms of a four-dimensional manifold. In the canonical formalism this simplicity is lost. The generators of the infinitesimal transformations do not even form a Lie-algebra, owing to the field dependent structure functions of their Poisson algebra. Nevertheless, nothing is lost from the local symmetries of solutions in the Lagrangian formulation as we pass to the Hamiltonian formalism in the following sense. Consider $\hat{\mathscr{F}}\xrightarrow{\,{}_{\mbox{$\scriptstyle\uppi$}}\;}\hat{\G}_\mathrm{ADM}$, where the bundle manifold $\hat{\mathscr{F}}$ consists of maximal globally hyperbolic vacuum spacetimes. $\hat{\G}_\mathrm{ADM}$ and $\mathrm{CAN}_\mathrm{ADM}$ are as before. See \refonline{lee_wald} for more details about the construction of these manifolds. Let $\mathrm{Diff}$ be the group of not necessarily metric independent spacetime diffeomorphisms. Then $\Uppi:\hat{\mathscr{F}}/\mathrm{Diff}\to\hat{\G}_\mathrm{ADM}/\mathrm{CAN}_{\mathrm{ADM}}$, $\,\Uppi(\mathrm{Diff}\cdot\mathbold{g})\define\mathrm{CAN}_\mathrm{ADM}\cdot\uppi(\mathbold{g})$  is a well-defined bijection.\cite{lee_wald,komar_bergmann} If we accept that plausible theories share this property with general relativity, the canonical formalism is an appropriate framework for our investigations, and we can use it without loss of generality. In our case the possible extensions of the spatial diffeomorphism algebra are yet unknown, so we do not know \emph{a priori} in what formalism, if any, the symmetry group takes such a simple form as the spacetime diffeomorphisms in the Lagrangian formalism of general relativity.

Similarly to Hojman, Kucha\v{r}, and Teitelboim,\cite{HKT,teitelboim,kuchar} we assume that the canonical variables are the spatial metric and its conjugate momentum, the same as in Einstein's gravity. The basic difference between their analysis and ours is that they assumed that the symmetry algebra is the canonical representation of the surface deformation algebra, which is the algebra of deformations of an $n$ dimensional spacelike surface embedded in an $n+1$ dimensional manifold, where the deformations are induced by the infinitesimal diffeomorphisms of the $n+1$ dimensional space. In their case the Poisson algebra of the generators was completely known. In our case the $n$ dimensional diffeomorphisms of the surface are part of the symmetry transformations, but unlike Hojman, Kucha\v{r}, and Teitelboim, we make no assumption on the form of the entire symmetry algebra. We are interested in the existence of algebras different from the representation of the surface deformation algebra rather than the uniqueness of a realization of a fully specified algebra, which is the subject of the analysis in \refonline{HKT,teitelboim,kuchar}. 

Section \ref{sec_preliminaries} and \ref{sec_supmom} contain some preparatory definitions and a discussion of the momentum constraint. Section \ref{sec_supham} analyzes the possibility of extending the supermomentum, which generates the spatial diffeomorphisms, by an additional generator function, whose form is taken to be a local kinetic term quadratic in the momentum $\pi^{i\j}$ together with a potential of quite general form. Under some simplifying assumptions on the structure functions, we find no extension of the spatial diffeomorphisms within this class, apart from the symmetry algebra of general relativity and its ultralocal truncation. Ho\v{r}ava's proposal belongs to this class.

In Section \ref{sec_uv} we modify Ho\v{r}ava's candidate for the UV fixed point theory. The phase space of this modification is reduced to the variables which are invariant under the local scale transformation. The constraints on the the phase space are first class, and they all generate symmetries on the solutions. In Section \ref{sec_disc} we mention some possibilities that have not been ruled out by the negative results of Section \ref{sec_supham}. 

Several appendices are attached to the paper. One of their purposes is to maintain the mathematical rigor of our analysis. Appendix \ref{assumptions_app} is devoted to some technical assumptions, which could be replaced by stronger locality conditions on the quantities arising in the Poisson algebra of the constraint functions. Nevertheless, this appendix helps us see what properties are really important for the final conclusion, and we can keep closer to the generality of \refonline{teitelboim}, where no locality assumption was made on the metric dependence of the elements of the Poisson algebra. In light of speculations that events might not play any essential role in the theory of gravity, the possibility of nonlocal gravitational potential that partially smears out events is worth considering. Appendices \ref{supmom_app}, \ref{elasticity_app}, and \ref{embeddability_app} describe some mathematical properties of the constraint functions, which the reader might find interesting in their own right, and which could be used in a further investigation of the symmetry algebra.

\section{Technical preliminaries and notations}
\label{sec_preliminaries}
\noindent Throughout the paper a symbol with an arrow on it denotes a vector field, a boldface letter stands for a general tensor density, or sometimes in the appendices, a tensor field valued linear map. Normalface letter with indices denotes the components of a tensor density or the tensor density itself, without indices, its trace (e.g. $\ppi$, $\pi^{i\j}$, $\pi$).     

In the Hamiltonian formulation of Einstein's gravity the canonical variables are the spatial metric $g_{i\j}$ and its conjugate momentum $\pi^{i\j}$. In our quest for a more general class of theories in which the temporal diffeomorphisms are replaced by some other symmetry the canonical variables are the same as in Einstein's gravity: $\f=(g_{i\j},\pi^{i\j})$.  The space $\Sigma$ is an $n$ dimensional manifold.

Let $\hh_\mu=(\hh,\hh_i)$ be a collection of functionals of the canonical variables. We will consider infinitesimal transformations whose generators, which are actually constraints, are $\sum_\mu\smallint_\Sigma\alpha^\mu\hh_\mu$, where $\alpha^\mu=(\alpha,\alpha^i)=(\alpha,\vec{\alpha})$ are the parameters of the transformation, also known as descriptors.\cite{komar_bergmann} The collection $\alpha^\mu$ is denoted by $\bar{\alpha}$ when too many indices would clutter up a formula. Traditionally, the super-Hamiltonian $\hh$ and the supermomentum $\hh_i$, called sometimes Hamiltonian and momentum constraints, are scalar and covector densities on $\Sigma$ of weight $1$. $\alpha$ is a scalar, $\alpha^i$ is a vector field. Coordinate independent notations would be cumbersome because of the presence of densities of nonzero weight. Any integral in the main text is meant with respect to a fixed coordinate volume element on $\Sigma$, and we represent densities $\tilde{T}$ by $\tilde{T}=\sqrt{g}T$, where $T$ is a tensor independent of the choice of the volume element. On the other hand, in the appendices any integration is meant with respect to the natural volume element. Since the pairing of $\hh_\mu$ with parameter $\alpha^\mu$ will appear frequently in our analysis, we will abbreviate it by
\[
\alpha^\mu\cd\hh_\mu\define\midsum_{\mu=0}^n\midint_\Sigma\alpha^\mu\hh_\mu.
\]
If $F$ is a functional of the canonical variables, its infinitesimal transformation generated by $\alpha^\mu$ is given by the Poisson bracket
\[
\delta_{\bar{\alpha}}F=\{F,\alpha^\mu\cd\hh_\mu\}.
\]
The parameter functions $\alpha^\mu$ and $\beta^\mu$ have to have some appropriately prescribed boundary values or decaying properties in order that the Poisson bracket $\{\alpha^\mu\cd\hh_\mu,\beta^\mu\cd\hh_\mu\}$, which is defined by functional derivatives, exist. We shall assume that it exists if both $\alpha^\mu$ and $\beta^\mu$ are in $\ccc$, independent of $(g_{i\j},\pi^{i\j})$. The symbol $\ccc$ denotes the space of compactly supported smooth functions or vector fields on $\Sigma$.

We give our definition of functional differentiability and some other properties of functionals. The argument of a general functional is enclosed by brackets, while the variable of an ordinary function is in parentheses.   

\begin{definition}
Let $\f$ be a collection of smooth tensor densities on the space $\Sigma$, $\G$ the manifold of configurations $\f$ on which the functionals $F:\G\to\mathbb{R}$ and $H$ are defined. $H[\f]$ is a tensor density on $\Sigma$.
\begin{itemize}
\item $F$ is {\it functionally differentiable} if for any variation, i.e.\ for any one-parameter family of $(\f_\lambda)_{\lambda\in(0,1)}\in\G$ for which $(\lambda,x)\mapsto\f_\lambda(x)$ is smooth and $\f_\lambda$ has appropriate boundary values or asymptotic properties, the derivative $\frac{\d}{\d\lambda}F[\f_\lambda]$ exists, and there is a collection $\mathcal{T}[\f_\lambda]$ of tensor densities on $\Sigma$ such that $(\lambda,x)\mapsto\mathcal{T}[\f_\lambda](x)$ is smooth, and
\[
\frac{\d F[\f_\lambda]}{\d\lambda}=\midsum_m\midint_\Sigma\mathcal{T}^m[\f_\lambda]\,\partial_\lambda{\f^m}_{\!\!\!\!\!\lambda\,\,},
\]
where the sum extends over the label $m$ of the members of the collection $\mathcal{T}$ and $\f$, and all the tensor indices of $\mathcal{T}^m$, which are suppressed, are contracted with the corresponding indices of $\f^m$. The derivative $\partial_\lambda\f$ is simply the partial derivative of the smooth function $(\lambda,x)\mapsto\f_\lambda(x)$ with respect to the first variable. $\mathcal{T}[\f]$ is called the functional derivative of $F$ at $\f$, and denoted by
\[
\mathcal{T}[\f](x)\define\frac{\delta F}{\delta\f(x)}.
\] 
\item $H$ is {\it local} in $\f$ if $\supp(H[\f_1]-H[\f_2])\subset\supp(\f_1-\f_2)$ for all $\f_{1,2}\in\G$.
\item $H$ is {\it ultralocal} in $\f$ if it is a function of $\f$ (but not its derivatives), that is, there is a function $h$ such that $H[\f](x)=h(\f(x))$ for all $x\in\Sigma$.
\item $H$ is {\it concomitant} of $\f$ if $H[f_*\f]=f_*H[\f]$ for any diffeomorphism $f:\Sigma\to\Sigma$.   
\end{itemize}
\end{definition}
The definition of ``concomitant'' gives a precise mathematical meaning to the property which is sometimes described as ``constructed out of $\f$ in a form invariant way'' or ``depends solely on $\f$''. Note the difficulty of giving a sensible definition to this property without assuming any tensorial structure on $H[\f]$. Hence the definition of ``concomitant'' includes the assumption of some tensorial structure so that the Lie transport of these quantities is defined.    

Unbarred $\alpha$ denotes not only the $0$th component of $\bar{\alpha}$, but if it is written in place of $\bar{\alpha}$ (as in $\delta_\alpha f$, $C^\mu[\alpha,\bar{\beta},Q]$, etc.) then it denotes an $\bar{\alpha}$ which is given by $\bar{\alpha}=(\alpha,\vec{0})$. The same convention applies to $\delta_{\vec{\alpha}}f$ etc., i.e., if $\vec{\alpha}$ stands in place of $\bar{\alpha}$, it refers to $\bar{\alpha}=(0,\vec{\alpha})$. 

We say that the Poisson algebra of $\hh_\mu$ closes if there are structure functions $C^\mu=(C,C^i)$ such that  
\eq{
\{\,\alpha^\mu\cd\hh_\mu\,,\beta^\mu\cd\hh_\mu\,\}=C^\mu[\bar{\alpha},\bar{\beta},\g,\ppi]\cd\hh_\mu,
\label{closure}
}
for any $\alpha^\mu,\beta^\mu\!\in\!\ccc$, independent of $(g_{i\j},\pi^{i\j})$, and a similar equality holds if they are obtained as structure functions in some previous Poisson bracket. We indicated that $C^\mu$ are functionals of the parameters and the canonical variables. $C^\mu[\bar{\alpha},\bar{\beta},\g,\ppi]$ are functions on $\Sigma$, but that is not the reason for their name ``structure function.'' This name refers to their dependence on the canonical variable, which prevents the Poisson algebra from having an ordinary Lie algebra structure, in which case the term ``structure constant'' would be appropriate. For further technical assumptions see Appendix \ref{assumptions_app}.

\section{Supermomentum}
\label{sec_supmom}
The supermomentum, $\hh_i$, is assumed to generate the spatial Lie transport of the canonical variables $\f=(g_{i\j},\pi^{i\j})$. Let $\vec{\alpha}$ be a vector field on the $n$ dimensional space $\Sigma$, independent of $\f$. Our assumption on $\hh_i$ is that\footnote{Strictly speaking, only $\mathbold{t}\cd\g=\smallint t^{i\j}g_{i\j}$ and $\mathbold{t}\cd\ppi=\smallint t_{i\j}\pi^{i\j}$ can have well-defined Poisson brackets with the generators $\ah$, where $\mathbold{t}$ is a tensor density. When we write $\{F[\f](x),\alpha^\mu\cd\hh_\mu\}=G[\bar{\alpha},\f](x)$, we mean that $G[\alpha^\mu,\f]$ is the smooth tensor density for which $\{\mathbold{t}\cdot F[\f],\alpha^\mu\cd\hh_\mu\}=\mathbold{t}\cdot G[\bar{\alpha},\f]$ holds for any compactly supported smooth tensor field $\mathbold{t}$ which is independent of $\f=(\g,\ppi)$. The  tensor indices of $\mathbold{t}$, $F$, and $G$ are suppressed.}
\eq{
\label{kinematic_spatial}
\mathcal{L}_{\vec{\alpha}}\f=\delta_{\vec{\alpha}}\f=\{\f,\alpha^i\cd\hh_i\}.
}
The Poisson bracket of two functionals $F$ and $G$ of $\f$ is defined as
\eq{
\label{standard_poisson}
\{F,G\}\define\frac{\delta F}{\delta g_{i\j}}\cd\frac{\delta G}{\delta\pi^{i\j}}-\frac{\delta G}{\delta g_{i\j}}\cd\frac{\delta F}{\delta\pi^{i\j}}
}
Using the Lie derivative of the metric and the momentum, which is a contravariant symmetric tensor density of weight 1, we get that $\hh_i$ satisfies  
\[
%\delta_{\vec{\alpha}}g_{i\j}
\{g_{i\j},\alpha^k\cd\hh_k\}=2\nabla_{(i}\alpha_{\j)}=\alpha^k\partial_k g_{i\j}+2g_{k(i}\partial_{\j)}\alpha^k,\;\;\;
%\delta_{\vec{\alpha}}\pi^{i\j}
\{\pi^{i\j},\alpha^k\cd\hh_k\}=\partial_k(\alpha^k\pi^{i\j})-2\pi^{k(i}\partial_k\alpha^{j)}.
\]
If $T$ is a tensor independent of the canonical variables, then the transformation of its covariant derivatives is
\[
\delta_{\vec{\alpha}}\nabla_{i_1}\dots\nabla_{i_m}T=\{\nabla_{i_1}\dots\nabla_{i_m}T,\alpha^i\cd\hh_i\}=\mathcal{L}_{\vec{\alpha}}\nabla_{i_1}\dots\nabla_{i_m}T-\nabla_{i_1}\dots\nabla_{i_m}\mathcal{L}_{\vec{\alpha}}T,
\]
where we suppressed the indices of $T$. $\nabla$ is the torsion free covariant derivative compatible with the metric $g_{i\j}$. Thus for $\vec{\alpha}$ and $\vec{\beta}$ independent of $(g_{i\j},\pi^{i\j})$:
\[
\delta_{\vec{\alpha}}\delta_{\vec{\beta}}g_{i\j}=\mathcal{L}_{\vec{\alpha}}\mathcal{L}_{\vec{\beta}}g_{i\j}-\mathcal{L}_{\mathcal{L}_{\vec{\alpha}}\vec{\beta}}g_{i\j}.
\]
$\mathcal{L}_{\vec{\alpha}}\mathcal{L}_{\vec{\beta}}g_{i\j}-\mathcal{L}_{\vec{\beta}}\mathcal{L}_{\vec{\alpha}}g_{i\j}=\mathcal{L}_{[\vec{\alpha},\vec{\beta}]}g_{i\j}$, and $\mathcal{L}_{\vec{\alpha}}\vec{\beta}=[\vec{\alpha},\vec{\beta}]$, hence
\eq{
\label{comm_spatial}
[\delta_{\vec{\alpha}},\delta_{\vec{\beta}}]g_{i\j}=-\mathcal{L}_{[\vec{\alpha},\vec{\beta}]}g_{i\j}=-\delta_{[\vec{\alpha},\vec{\beta}]}g_{i\j}.
}
For any spatial metric we have
\eq{
\label{supmom_poisson_derivation}
\begin{split}
\{g_{i\j},[\vec{\alpha},\vec{\beta}]^k\cd\hh_k\}&=\delta_{[\vec{\alpha},\vec{\beta}]}g_{i\j}=-[\delta_{\vec{\alpha}},\delta_{\vec{\beta}}]g_{i\j}=-(\delta_{\vec{\alpha}}\delta_{\vec{\beta}}-\delta_{\vec{\beta}}\delta_{\vec{\alpha}})g_{i\j}\\&=-\{\{g_{i\j},\beta^k\cd\hh_k\},\alpha^l\cd\hh_l\}+\{\{g_{i\j},\alpha^k\cd\hh_k\},\beta^l\cd\hh_l\}=\{g_{i\j},\{\alpha^k\cd\hh_k,\beta^l\cd\hh_l\}\}
\end{split}
}
where in the last equality the Jacobi identity was used. Therefore equation \eqref{comm_spatial} already fixes the Poisson algebra of $\hh_i$:
\eq{
\label{supmom_poisson}
\{\alpha^i\cd\hh_i,\beta^j\cd\hh_j\}=[\vec{\alpha},\vec{\beta}]^i\cd\hh_i
}
Note that if $\vec{\alpha}$ and $\vec{\beta}$ are independent of the canonical variables, so is $[\vec{\alpha},\vec{\beta}]$, hence \eqref{kinematic_spatial} was indeed applicable in the first equality in \eqref{supmom_poisson_derivation}. We quote the following result from \refonline{HKT}:
\begin{prop}
If $\hh_i$ is concomitant of the canonical variables $(g_{i\j},\pi^{i\j})$, and $\mathcal{L}_{\vec{\alpha}}g_{i\j}=\{g_{i\j},\alpha^k\cd\hh_k\}$, $\mathcal{L}_{\vec{\alpha}}\pi^{i\j}=\{\pi^{i\j},\alpha^k\cd\hh_k\}$ for any $\alpha^i,\beta^i\in\ccc$, independent of $(g_{i\j},\pi^{i\j})$, then the only possible $\hh_i$ are 
\eq{
\label{eq_supmom}
\hh_i=-2g_{i\j}\sqrt{g}\nabla_k\frac{\pi^{jk}}{\sqrt{g}}.
}
\end{prop}
In section \ref{sec_uv} we will define a dynamics on conformal classes of metrics. To this end, it will be useful to realize the spatial diffeomorphism algebra on the reduced phase space of conformal classes of metrics.  

Let $f:\Sigma\to\Sigma$ be a diffeomorphism. A metric $g_{i\j}$ and $g'_{i\j}$ are said to be in the same conformal class if there is a function $\sigma$ on $\Sigma$ such that $g'_{i\j}=g_{i\j}e^\sigma$. For a generic $f$ the metrics $f_*g_{i\j}$ and $f_*g'_{i\j}$ are not in the same conformal class even if $g_{i\j}$ and $g'_{i\j}$ are. However, it is possible to define the transformation of the metric so that it is a map between conformal classes.
Define 
\eq{
\label{supmom_conf_def}
\tilde{\hh}_i\define\hh_i+\frac{2}{n}\sqrt{g}\nabla_i\frac{\pi}{\sqrt{g}},
} 
where $\pi=\pi^i_i$. The infinitesimal transformation generated by $\alpha^i\cd\tilde{\hh}_i$ on the metric is
\[
%\label{poisson_supmom_conf}
\{g_{i\j},\alpha^k\cd\tilde{\hh}_i\}=2\nabla_{(i}\alpha_{\j)}-\frac{2}{n}g_{i\j}\nabla_k\alpha^k,
%(P\vec{\xi})_{i\j},
%\quad\quad\quad\{\xi^i\cd\tilde{\hh}_i,\eta^j\cd\tilde{\hh}_j\}=[\vec{\xi},\vec{\eta}]^i\cd\tilde{\hh}_i,
\]
The Poisson algebra of $\tilde{\hh}_i$ is the same as that of $\hh_i$, given by \eqref{supmom_poisson}. Let us introduce the conformal metric and momentum:\cite{HRT}
\eq{
\label{conf_var}
\tilde{g}_{i\j}\coloneqq g^{-\frac{1}{n}}g_{i\j},\;\;\;\;\;\tilde{\pi}^{i\j}=g^\frac{1}{n}\left(\pi^{i\j}-\frac{1}{n}g^{i\j}\pi\right),
}
The conformal supermomentum \eqref{supmom_conf_def} is an expression of the conformal variables \eqref{conf_var} only:
\eq{
\tilde{\hh}_i=-2\tilde{g}_{i\j}\tilde{\nabla}_k\tilde{\pi}^{jk},
\label{eq_supmom_conf}
}
where $\tilde{\nabla}$ is the covariant derivative compatible with the metric $\tilde{g}_{i\j}$. The transformation of the conformal variables generated by $\alpha^i\cd\tilde{\hh}_i$:
\[
%\label{conf_transform}
\{\tilde{g}_{i\j},\alpha^k\cd\tilde{\hh}_k\}=2\tilde{\nabla}_{(i}\alpha_{\j)}-\frac{2}{n}\tilde{g}_{i\j}\tilde{\nabla}_k\alpha^k,\;\;\;\{\tilde{\pi}^{i\j},\alpha^k\cd\tilde{\hh}_k\}=\partial_k(\alpha^k\tilde{\pi}^{i\j})-2\tilde{\pi}^{k(i}\partial_k\alpha^{j)}+\frac{2}{n}\tilde{\pi}^{i\j}\,\tilde{\nabla}_k\alpha^k.
\]

\section{Super-Hamiltonian}
\label{sec_supham}
The question that we turn to now is what are the possible super-Hamiltonians $\hh$ such that the Poisson algebra of $\hh_\mu$ closes in the sense \eqref{closure}, where $\hh_i$ are given by \eqref{eq_supmom}.  
We shall consider super-Hamiltonians $\hh$ which are (i) ultralocal and (ii) at most quadratic in the momentum. We also assume that (iii) $\hh$ does not contain a term linear in the momentum $\pi^{i\j}$, so $\hh$ is the sum of a momentum independent potential term and a kinetic term which is a homogeneous quadratic expression of $\pi^{i\j}$. Finally, we assume that (iv) the kinetic term is ultralocal in the metric.  

In \refonline{HKT} condition (i) is the consequence of a kinematical condition: The transformation of the spatial metric generated by $\alpha\cd\hh$ is required to be ultralocal in $\alpha$ as in general relativity. Hojman, Kucha\v{r}, and Teitelboim were interested in the possibility of different Einsteinian geometrodynamics, i.e.\ the existence of $\hh$ such that $(\hh,\hh_i)$ give inequivalent canonical realizations of the symmetry algebra of Einstein's gravity.\cite{teitelboim,HKT,kuchar} If (ii) is also assumed, then Einstein's gravity is the only realization of that algebra even for spatial dimensions higher than three.\cite{teitelboim} In this case (iii) and (iv) are consequences of the symmetry algebra. It was argued in \refonline{HKT,kuchar} that if the space is three dimensional, then even (ii) can be relaxed: The kinetic term is allowed to be a power series of the momentum with metric dependent coefficients. In \refonline{HKT} only time reversible geometrodynamics is considered, i.e., $\hh$ is taken to be an even function of the momentum, but this condition turns out to be redundant.\cite{kuchar} Thus for a three dimensional space (i) is enough to regain Einstein's gravity as the only Einsteinian geometrodynamics. 

Since in our case the symmetry algebra is only partially known, the analysis for such a general class of $\hh$ as in Hojman, Kucha\v{r}, and Teitelboim's investigations would be much more complicated than the exploration of the different realizations of the fixed symmetry algebra of Einstein's gravity. Dropping (i) would probably make the analysis of Hojman {\it et al} intractable, and in our case it would allow for uninteresting modifications of Einstein's gravity. For example $\hh_\mu\!=\!(\hh_\mathrm{ADM}\!+\!t^{i\j}\hh_i\hh_j,\,\hh_i)$, where $\hh_\mathrm{ADM}$ is the Hamiltonian constraint of general relativity, and $t^{i\j}$ is a tensor density of weight $-1$, concomitant of the canonical variables. Theories with $\hh$ satisfying (ii) appear to be the physically most relevant ones since the relationship between the momentum and the velocity is linear, so the momentum can be uniquely eliminated in favor of the velocity if the kinetic term is a nondegenerate expression of the momentum. Condition (iii) can be interpreted as the requirement of time reversibility.\cite{HKT} For assumption (iv) we do not have any physical motivation. It is assumed for the sake of simplicity of our analysis, but we note that Ho\v{r}ava's proposal and its later modifications, including the ones in which the detailed balance condition is broken, all belong to the class to which the theorem in this section applies.

The question of different extensions of spatial diffeomorphisms was already raised in \refonline{teitelboim}, where Teitelboim notes that we do not know what makes the Poisson algebra of general relativity preferable over other possibilities, and he mentions ultralocal gravity as an alternative to Einstein's gravity. As far as we know, all the attempts to find theories which admit an extension of the spatial diffeomorphisms with a further local symmetry, and in which $(g_{i\j},\pi^{i\j})$ are the only canonical variables, satisfy condition (i-iv). In \refonline{BFM} the following potentials terms in $\hh$ were tried: $V\!=\!\mu R\!+\!\nu$, $V\!=\!R^a$; and the potential $V\!=\!c_1R^2\!+\!c_2R_{i\j}R^{i\j}\!+\!c_3\nabla_i\nabla^iR$ is also claimed to have been tested ($\mu,\nu,a,c_1,c_2,c_3$ are constants, $R_{i\j}$ is the Ricci tensor of the spatial metric $g_{i\j}$, $R\!=\!R_{i\j}g^{i\j}$). None of them were found to result in a closed Poisson algebra of $\hh_\mu$. In \refonline{kocharyan} it is conjectured that if the kinetic term in $\hh$ is the same as in the Hamiltonian constraint of general relativity, then the Poisson algebra of $\hh_\mu$ closes only for Einstein's gravity. Apart from (i-iv), some mathematical properties described in Appendix \ref{assumptions_app}, and a simplifying condition on the momentum dependence of the structure functions that we will describe soon, we do not make further assumptions. In particular, we do not use some specific form for the potential, but it is allowed to be a general, possibly nonlocal functional of the metric. So our analysis applies to $\hh$ of the form $\hh=\frac{1}{\sqrt{g}}G_{i\j kl}\pi^{i\j}\pi^{kl}+G[\g]$, where $G_{i\j kl}$ is ultralocal in the spatial metric $g_{i\j}$, but $G[\g]$ can be a general functional of $g_{i\j}$.            

In Einstein's gravity the structure functions $C^i[\alpha,\beta,\g]$ in \eqref{closure} depend on the metric. In general $C^\mu[\bar{\alpha},\bar{\beta},\g,\ppi]$ can depend both on the metric $g_{i\j}$ and the momentum $\pi^{i\j}$. However, we will consider only the case when $C[\alpha,\beta,\g,\ppi]$ is at most linear in the momentum. Recall our notations: This means only an assumption on the structure function with which $\hh$ might arise on the right hand side of its own Poisson brackets. In the theorem below, for clarity, any possible momentum and metric dependence is indicated explicitly.
\begin{theor}
\label{linC}
Let the space $\Sigma$ be $n\geq3$ dimensional. $\hh_i[\g,\ppi]$ is the supermomentum given by \eqref{eq_supmom}, and $\hh[\g,\ppi]=\frac{1}{\sqrt{g}}\,G_{i\j kl}\pi^{i\j}\pi^{kl}+G[\g]$, where $G_{i\j kl}=\kappa\,(g_{ik}g_{jl}+g_{il}g_{jk})-\lambda\,g_{i\j}g_{kl}$ with constants satisfying $n\lambda\neq 2\kappa\neq 0$. Assume that for any $\alpha,\beta\in\ccc$ there is a smooth tensor field $\mathbold{t}[\alpha,\beta,\g]$ such that $\supp\mathbold{t}[\alpha,\beta,\g]\subset\supp\alpha\cup\supp\beta$, and for any momentum $\pi^{i\j}$
\[
%\label{C_prop}
C[\alpha,\beta,\g,\ppi]\cd\hh[\g,\ppi]+C^i[\alpha,\beta,\g,\ppi]\cd\hh_i[\g,\ppi]=\midint_\Sigma t_{i\j}[\alpha,\beta,\g]\pi^{i\j}
\]
with smooth functions $C[\alpha,\beta,\g,\ppi]$ and $C^i[\alpha,\beta,\g,\ppi]$, which are linear in $\alpha$ and $\beta$, and their support is within $\supp\alpha\cup\supp\beta$. Furthermore, $C[\alpha,\beta,\g,\ppi]$ is also linear in $\pi^{i\j}$. Then there are smooth vector fields $\vec{v}[\alpha,\beta,\g]$, linear in $\alpha$ and $\beta$, such that $\supp\vec{v}[\alpha,\beta,\g]\subset\supp\alpha\cup\supp\beta$, and for any momentum $\pi^{i\j}$ 
\[
%\label{reexpress}
\midint_\Sigma t_{i\j}[\alpha,\beta,\g]\pi^{i\j}=\midint_\Sigma v^i[\alpha,\beta,\g]\hh_i[\g,\ppi].
\]
\end{theor}

In other words, under the simplifying assumptions on the structure functions mentioned in the above theorem, if the Poisson bracket is a linear expression of the momentum, which is true if $\hh$ is of the form as in the above theorem, and it closes, then it is a linear expression of the supermomentum itself. So the structure function $C[\alpha,\beta,\g,\ppi]$ actually vanishes. This statement is plausible, but the rigorous proof, which can be found in Appendix \ref{proof_app}, is somewhat technical, in part owing to the relatively weak locality assumptions on $C^\mu[\alpha,\beta,\g,\ppi]$. Actually, the statement is plausible even if we relax the condition that $C[\alpha,\beta,\g,\ppi]$ is linear in $\pi^{i\j}$, especially if we assume that $C^\mu[\alpha,\beta,\g,\ppi](x)$ is a polynomial of $\pi^{i\j}$ and its derivatives at $x\in\Sigma$. If the degree of the momentum dependence of the structure function accompanying $\hh$ is as high as in $\hh$ itself, the appeal of the canonical formalism seems to be lost. We are not going to extend our analysis in this direction. The following theorem applies to the case when the Poisson bracket $\{\alpha\cd\hh,\beta\cd\hh\}$ closes on the supermomentum, i.e., $C[\alpha,\beta,\g,\ppi]=0$, which is true if $C[\alpha,\beta,\g,\ppi]$ is required to be independent of the momentum, or if it depends linearly on the momentum, and $C^\mu$ satisfy the conditions of the previous theorem. In the next theorem the metric and momentum dependence of $\hh_\mu$ and the structure functions will not be indicated explicitly. The parameters $\alpha^\mu,\beta^\mu$ are always assumed to be independent of the canonical variables. 
\begin{theor}
\label{main}
Assume that the Poisson algebra of $\hh_\mu=(\hh,\hh_i)$ closes with the standard Poisson bracket \eqref{standard_poisson} on $\hh_i$. $\hh_\mu$ are functions on the $n\geq2$ dimensional space $\Sigma$, concomitant of the canonical variables $(g_{i\j},\pi^{i\j})$, and  
\[
\begin{split}
&\hh=\frac{1}{\sqrt{g}}G_{i\j kl}\pi^{i\j}\pi^{kl}+G[\g],\\&\mathcal{L}_{\vec{\alpha}}g_{i\j}=\{g_{i\j},\alpha^i\cd\hh_i\},\;\;\;\mathcal{L}_{\vec{\alpha}}\pi^{i\j}=\{\pi^{i\j},\alpha^i\cd\hh_i\}\quad\mbox{for any}\quad\alpha^i,\beta^i\in \ccc,
\end{split}
\]
where $G_{i\j kl}$ is an ultralocal, but $G[\g]$ a general (not necessarily local) functional of the metric. Under these conditions $\hh$ can always be rescaled so that the Poisson algebra of $\hh_\mu$ is
\[
\begin{split}
&\{\alpha\cd\hh,\beta\cd\hh\}=-\epsilon g^{i\j}(\alpha\,\partial_{\!j}\beta-\beta\,\partial_{\!j}\alpha)\cd\hh_i,\\&\{\alpha^i\cd\hh_i,\beta\cd\hh\}=(\alpha^i\partial_i\beta)\cd\hh,\\&\{\alpha^i\cd\hh_i,\beta^j\cd\hh_j\}=[\vec{\alpha},\vec{\beta}]^i\cd\hh_i,
\end{split}
\]
where $\epsilon=0$ (symmetry algebra of ultralocal gravity) or $\epsilon=\pm 1$ (symmetry algebra of Einstein's gravity with Euclidean ($\epsilon=1$) or Lorentzian ($\epsilon=-1$) signature). 

If $G_{i\j kl}$ is an invertible map between the spaces of rank two symmetric tensors, then $G[\g]$ is a function of the metric and its first and second derivatives. If $\epsilon=0$ also holds, then $G[\g]$ is actually ultralocal in the metric. If $\epsilon\neq0$, then
$G_{i\j kl}\!=\!\kappa\,(g_{ik}g_{\j l}+g_{il}g_{\j k}-\frac{2}{n-1}\,g_{i\j}g_{kl})$, where $\kappa\neq0$ is constant.

If $G_{i\j kl}$ is not invertible, then $\epsilon=0$.
\end{theor}
\begin{proof}
The proposition in Section \ref{sec_supmom} has already established the only possible form of $\hh_i$. We divide the analysis of $\hh$ into five steps. See Section \ref{sec_preliminaries} for our notational conventions.
\vspace{1ex}

\noindent{\it 1. Auxiliary noncanonical variables}

\vspace{1ex}
\noindent 
Since $G_{i\j kl}$ is assumed to be ultralocal in the metric, the Poisson brackets of $\alpha\cd\hh$ and $\beta\cd\hh$ can produce only linear functionals of the momentum. By the assumption $C[\alpha,\beta]=0$, these Poisson brackets must close on the supermomentum $\hh_i$ with structure functions $C^i[\alpha,\beta]$ independent of the momentum. $\hh$ is assumed to be a scalar density, concomitant of the canonical variables, the Poisson bracket of $\hh$ with $\beta^i\cd\hh_i$ is the Lie derivative of $\hh$ with respect to $\vec{\beta}$, which is a linear homogeneous expression of $\hh$ and its first derivatives. $C^i[\alpha,\vec{\beta}]=0$, and the weight of $\hh$ fixes $C[\alpha,\vec{\beta}]$. What will be important first is that $C^\mu[\alpha,\bar{\beta}]$ are all independent of the momentum. Finally, the structure functions $C^i[\vec{\alpha},\vec{\beta}]=[\vec{\alpha},\vec{\beta}]^i$, which are independent of the canonical variables, and $C[\vec{\alpha},\vec{\beta}]=0$ are already known from \eqref{supmom_poisson}. Note that $C^\mu[\bar{\alpha},\bar{\beta}]$ is bilinear.

Let us introduce the auxiliary variables $\bar{\nn}\!=\!\nn^\mu\!=\!(\nn,\nn^i)$, which is a collection of a scalar and a vector field in $\ccc$, independent of $g_{i\j}$ and $\pi^{i\j}$. By the assumption on the momentum dependence of $\hh$,
\[
%\label{defP}
\frac{\delta(\nn\cd\hh)}{\delta\pi^{i\j}(x)}=2\nn(x)P_{i\j}(x),\quad\mbox{where}\quad P_{i\j}\define\frac{1}{2}\frac{\partial\hh}{\partial\pi^{i\j}}.
\]
We introduced the quantity $P_{i\j}$ for the sake of brevity of later calculations. Note that
\eq{
\label{ghn}
\{g_{i\j},\nh\}=2\nn P_{i\j}+2\nabla_{(i}\nn_{\j)}.
}
By \eqref{ghn} the infinitesimal transformation of $g_{i\j}$ generated by $\beta\cd\hh$ satisfies
\eq{
\label{deltag_P}
\nn\delta_\beta g_{i\j}=\nn\{g_{i\j},\beta\cd\hh\}=2\nn\beta P_{i\j}=\beta\left(\{g_{i\j},\nh\}-2\nabla_{(i}\nn_{\j)}\right).
}
In order to get a similar expression for the second variation $\delta_\beta\delta_\alpha g_{i\j}$ corresponding to two infinitesimal transformations, we have to exchange the order of the transformation and the Poisson-bracketing with $\nh$. The Jacobi identity tells us that
\[
%\label{exch_tr_timeev1}
\begin{split}
\left\{\{g_{i\j},\nh\},\ah\right\}&=\left\{\{g_{i\j},\ah\},\nh\right\}-\left\{g_{i\j},\{\ah\,\nh\}\right\}\\&=\left\{\{g_{i\j},\ah\},\nh\right\}-\left\{g_{i\j},C^\mu[\bar{\alpha},\bar{\nn}]\cd\hh_\mu\right\}.
\end{split}
\]
Using that $C^\mu$ are independent of the momentum, and $\alpha$ does not depend on the canonical variables, we can evaluate the right hand side:
\eq{
\label{exch_tr_timeev2}
\left\{\{g_{i\j},\nh\},\alpha\cd\hh\right\}=2\alpha\{P_{i\j},\nh\}-2\left(\nabla_{(i}C_{\j)}[\alpha,\bar{\nn}]+C[\alpha,\bar{\nn}]P_{i\j}\right).
}
From \eqref{deltag_P} and \eqref{exch_tr_timeev2} the second variation of the metric corresponding to two consecutive transformations generated by $\alpha\cd\hh$ and $\beta\cd\hh$, respectively:
\eq{
\label{secondvar}
\begin{split}
\nn\delta_\alpha\delta_\beta g_{i\j}&=\{\nn\delta_\beta g_{i\j},\alpha\cd\hh\}\\&=2\alpha\beta\{P_{i\j},\nh\}-2\beta\left(\nabla_{(i}C_{\j)}[\alpha,\bar{\nn}]+C[\alpha,\bar{\nn}]P_{i\j}+2\alpha P_{k(i}\nabla_{\j)}\nn^k+\nn^k\nabla_k(\alpha P_{i\j})\right),
\end{split}
}
where the last two terms came from the evaluation of the Poisson bracket of $2\nabla_{(i}\nn_{\j)}=2g_{k(i}\partial_{\j)}\nn^k+\nn^k\partial_kg_{i\j}$ with $\beta\cd\hh$. The commutator of the two transformations:
\begin{align}
\nn[\delta_\alpha,\delta_\beta]g_{i\j}&=\nn(\delta_\alpha\delta_\beta g_{i\j}-\delta_\beta\delta_\alpha g_{i\j})\nonumber\\
\label{HamHam}
&=2P_{i\j}\left(\alpha\left(C[\beta,\bar{\nn}]+\nn^k\partial_k\beta\right)-\beta\left(C[\alpha,\bar{\nn}]+\nn^k\partial_k\alpha\right)\right)\\
&\quad\quad\quad\quad\quad\quad\quad+2\left(\alpha\nabla_{(i}C_{\j)}[\beta,\bar{\nn}]-\beta\nabla_{(i}C_{\j)}[\alpha,\bar{\nn}]\right).\nonumber\\
\intertext{Similar calculations lead to the following expression of the commutator of two transformations generated by $\alpha\cd\hh$ and $\beta^i\cd\hh_i$, respectively:}
\label{HamMom}
\nn[\delta_\alpha,\delta_{\vec{\beta}}]g_{i\j}&=2\nn P_{i\j}\,\beta^k\partial_k\alpha+2\alpha\left(P_{i\j}\left(C[\vec{\beta},\bar{\nn}]-\beta^k\partial_k\nn\right)+g_{k(i}\nabla_{\j)}\left(C^k[\vec{\beta},\bar{\nn}]-\mathcal{L}_{\vec{\beta}}\nn^k\right)\right).
\end{align}

\noindent{\it 2. A note on the tensorial structure of $\hh$}
 
\vspace{1ex}
\noindent $\hh$ is a scalar density of weight $1$ by assumption. Multiplying $\hh$ by a power of $\sqrt{g}$, we get an $\hh$ of different weight. This freedom in changing the weight of $\hh$ is reflected by \eqref{HamHam} and \eqref{HamMom}. $C^i[\vec{\beta},\bar{\nn}]\!=\!C^i[\vec{\beta},\vec{\nn}]+C^i[\vec{\beta},\nn]$. We have already seen that $C^i[\vec{\beta},\nn]\!=\!0$. The last term in \eqref{HamMom} vanishes since the structure constants in the Poisson algebra of $\hh_i$ are precisely $C^i[\vec{\beta},\vec{\nn}]\!=\!\mathcal{L}_{\vec{\beta}}\nn^i$ (see \eqref{supmom_poisson}). We conclude from \eqref{HamMom} that 
\eq{
\label{C}
C[\vec{\beta},\bar{\nn}]=\beta^i\partial_i\nn+\nn\tilde{C}[\vec{\beta}],
}
where $\tilde{C}$ is possibly a functional of the metric, and a linear functional of $\beta^i$. Note that the term proportional to $P_{i\j}$ in \eqref{HamHam} completely cancels since $C[\beta,\nn]\!=\!0$, and $C[\beta,\bar{\nn}]\!=\!C[\beta,\vec{\nn}]\!=\!-\nn^i\partial_i\beta-\beta\tilde{C}[\vec{\nn}]$ from \eqref{C} and the antisymmetry of $C$ in the parameters. The Lie derivative of a scalar density $\hh$ of weight $w$ is $\mathcal{L}_{\vec{\beta}}\hh\!=\!\partial_i(\beta^i\hh)+(w-1)\hh\partial_i\beta^i$. Since $\mathcal{L}_{\vec{\beta}}\hh\!=\!\{\hh,\beta^i\cd\hh_i\}$ and 
\[
C[\vec{\beta},\nn]\cd\hh\!=\!C[\vec{\beta},\nn]\cd\hh+C^i[\vec{\beta},\nn]\cd\hh_i\!=\!-\{\nn\cd\hh,\beta^i\cd\hh_i\}\!=\!-\nn\cd\mathcal{L}_{\vec{\beta}}\hh\!=\!(\beta^i\partial_i\nn-\nn(w-1)\partial_i\beta^i)\cd\hh,
\] 
a scalar density $\hh$ of weight $w$ corresponds to $\tilde{C}[\vec{\beta}]=-(w-1)\partial_i\beta^i$. With our choice of weight $1$,
\eq{
\label{Cfinal1}
C[\vec{\alpha},\beta]=\alpha^i\partial_i\beta.
}

\noindent{\it 3. Locality}
 
\vspace{1ex}
\noindent Even if we allowed $G[\g]$ to be a nonlocal functional of the metric, we can rule out the possibility that the structure functions are nonlocal in the parameters. As mentioned in the previous paragraph, \eqref{HamHam} simplifies to 
\eq{
\label{HHsimpler}
\nn[\delta_\alpha,\delta_\beta]g_{i\j}=2\left(\alpha\nabla_{(i}C_{\j)}[\beta,\bar{\nn}]-\beta\,\nabla_{(i}C_{\j)}[\alpha,\bar{\nn}]\right).
}
Also, $C^i[\beta,\bar{\nn}]=C^i[\beta,\nn]$ because $C^i[\beta,\vec{\nn}]=0$. From \eqref{HHsimpler} we get
\eq{
\label{CC}
\alpha\nabla_{(i}C_{\j)}[\beta,\nn]|_{(\supp\beta)^c}=\nn\Phi_{i\j}[\alpha,\beta]|_{(\supp\beta)^c},
}
where the superscript ${}^c$ indicates complement. $\Phi_{i\j}$ (as $C^i$) can depend on the metric. Since $\alpha\!\in\!\ccc$ can be arbitrary,  
\[
\supp\nabla_{(i}C_{\j)}[\beta,\nn]|_{(\supp\beta)^c}\subset\supp\nn|_{(\supp\beta)^c},
\]
Since the only solution to the Killing equation on an open subset of $\Sigma$ is zero for a generic metric, we have
\[
\supp C_i[\beta,\nn]|_{(\supp\beta)^c}\subset\supp\nn|_{(\supp\beta)^c}. 
\]
A set of metrics is not generic if a structure function that is nonzero only on this set could not arise in an element of the Poisson algebra since the latter would not be functionally differentiable. By property \eqref{cont_structfunc_prop} in Appendix \ref{assumptions_app} and lemma \ref{lem_local}, there is a finite expansion for any $x\notin\supp\beta$ in terms of the derivatives of $\nn$ so that it gives $C_i[\beta,\nn]$ and its first derivatives at $x$. With this expansion in \eqref{CC}, we have
\[
\alpha\nabla^{\phantom{k}_{\phantom{1}}}_{(i}\sum_m^MC_{\j)}^{k_1k_2\dots k_m}(x)\nabla^{\phantom{k}_{\phantom{1}}}_{(k_1}\nabla^{\phantom{k}_{\phantom{1}}}_{k_2}\dots\nabla^{\phantom{k}_{\phantom{1}}}_{k_m)}\nn|_{(\supp\beta)^c}=\nn\Phi^{\phantom{k}_{\phantom{1}}}_{i\j}[\alpha,\beta]|_{(\supp\beta)^c},%\quad\mbox{if}\quad x\notin\supp\beta,
\]
where the coefficients $C_i^{k_1k_2\dots k_m}$ can depend on $\beta$ and the metric, and the number $M$ of the terms in the sum may vary with $x\in\S$. Since we can always specify $\nn$ so that precisely one of its symmetric covariant derivatives of the highest order appearing on the left hand side is nonzero, and all the others, including $\nn$ itself, are zero at a given point, the only possibility is that all the coefficients $C_i^{k_1k_2\dots k_m}(x)=0$ at $x\notin\supp\beta$. Thus $\supp C_i[\beta,\nn]\subset\supp\beta$. By property \eqref{cont_structfunc_prop} and lemma \ref{lem_local} again, this means that at any point in the entire space $\Sigma$ there is a finite combination of the derivatives of $\beta$ whose derivatives below a fixed order give $C_i$ and its corresponding derivatives at that point. The coefficient functions can depend on the metric and the other parameter $\nn$ as well. But $C_i$ is antisymmetric in the two parameters, so this expansion can be written in terms of the derivatives of the two parameters. So on the entire space $\Sigma$ we have
\eq{
\label{Cloc}
\nabla_{(i}^{\phantom{k}_{\phantom{1}}}C_{j)}^{\phantom{k}_{\phantom{1}}}[\beta,\nn]=\nabla_{(i}^{\phantom{k}_{\phantom{1}}}\!\!\sum_{m,n=0}^M\!\!C_{j)}^{k_1\dots k_m,l_1\dots l_n}\left(\nabla^{\phantom{k}_{\phantom{1}}}_{(k_1}\dots\nabla^{\phantom{k}_{\phantom{1}}}_{k_m)}\,\beta\right)\nabla^{\phantom{k}_{\phantom{1}}}_{(l_1}\dots\nabla^{\phantom{k}_{\phantom{1}}}_{l_n)}\nn
}
with possibly metric dependent coefficients $C_i^{k_1\dots k_m,l_1\dots l_n}$, which are totally symmetric in $k_1\dots k_m$ as well as in $l_1\dots l_n$, and $C_i^{k_1\dots k_m,l_1\dots l_n}=-C_i^{l_1\dots l_n,k_1\dots k_m}$. Note that $M$ can depend on $x\in\S$.
\vspace{1ex}

\noindent{\it 4. Structure functions}

\vspace{1ex}
\noindent Suppose that at some $x$ higher than first derivatives arise in \eqref{Cloc}. Let $m(x)$ be the maximum number of $k$-indices that occur in \eqref{Cloc}, and $n\in\mathbb{N}$. Fix the following two sets of indices: $K\define\{i,k_1,\dots,k_{m(x)}\}$ and $L\define\{l_1,\dots l_n\}$. Take an $\alpha$ which is a nonzero constant in a neighborhood of $x$, a $\beta$ for which $\beta(x)=0$ and $\partial_{k_1'}\dots\partial_{k_{m'}'}\beta(x)=0$ unless $\{k_1',\dots,k_{m'}'\}=K$, and an $\nn$ such that $\nn(x)=0$ and $\partial_{l_1'}\dots\partial_{l_{n'}'}\nn(x)=0$ unless $\{l_1',\dots,l_{n'}'\}=L$. What survives from the $ii$-component of \eqref{HHsimpler} with this choice is
\[
0=\nn(x)[\delta_\alpha,\delta_\beta]g^{\phantom{k}_{\phantom{1}}}_{ii}\!(x)=2\alpha(x)\;C_i^{k_1\dots k_{m(x)},l_1\dots l_n}(x)\left(\partial^{\phantom{k}_{\phantom{1}}}_i\!\partial^{\phantom{k}_{\phantom{1}}}_{k_1}\!\dots\partial^{\phantom{k}_{\phantom{1}}}_{k_{m(x)}}\,\beta(x)\right)\partial^{\phantom{k}_{\phantom{1}}}_{l_1}\dots\partial^{\phantom{k}_{\phantom{1}}}_{l_n}\nn(x),
\]
where there is no summation over $k_1\dots k_{m(x)}$ and $l_1\dots l_n$. This can hold for all such choices only if $n=0$, i.e., highest derivatives of $\beta$ in $C_i[\beta,\nn]$ is  multiplied by only undifferentiated $\nn$. By the antisymmetry of $C_i[\beta,\nn]$ in the parameters,
\eq{
\label{CHighestDer}
C^{\phantom{k}_{\phantom{1}}}_i\![\beta,\nn](x)=C_i^{k_1\dots k_{m(x)}}(x)\left(\nn(x)\,\partial^{\phantom{k}_{\phantom{1}}}_{k_1}\!\dots\partial^{\phantom{k}_{\phantom{1}}}_{k_{m(x)}}\,\beta(x)-\beta(x)\,\partial^{\phantom{k}_{\phantom{1}}}_{k_1}\!\dots\partial^{\phantom{k}_{\phantom{1}}}_{k_{m(x)}}\nn(x)\right)+\dots,
}
where the ellipsis indicates terms containing derivatives of $\beta$ and $\nn$ of order lower than $m(x)$. Now suppose that $m(x)\geq 2$. If we choose an $\nn$ such that $\nn(x)=0$, and its only derivative that does not vanish at $x$ is of order $m(x)$, and $\beta$ is a function such that only $\partial_i\beta(x)\neq0$ for a fixed $i$, and all its other derivatives, including $\beta$ itself, are zero at $x$, then from \eqref{HHsimpler} together with \eqref{CHighestDer} we get
\[
0=\nn(x)[\delta_\alpha,\delta_\beta]g^{\phantom{k}_{\phantom{1}}}_{ii}\!(x)=-2C_i^{k_1\dots k_{m(x)}}(x)\;\partial^{\phantom{k}_{\phantom{1}}}_i\!\beta(x)\;\partial^{\phantom{k}_{\phantom{1}}}_{k_1}\!\dots\partial^{\phantom{k}_{\phantom{1}}}_{k_{m(x)}}\nn(x),
\]
where there is no summation over $k_1\dots k_{m(x)}$. The realization is that it is impossible to satisfy the above equality unless $m(x)\leq 1$ for all $x$. So the possible structure functions reduce to
\eq{
\label{Cfinal2}
C^{\phantom{k}_{\phantom{1}}}_i\![\alpha,\beta]=C_i^j(\alpha\,\partial^{\phantom{k}_{\phantom{1}}}_{\!j}\!\!\beta-\beta\;\partial^{\phantom{k}_{\phantom{1}}}_{\!j}\!\!\alpha)
}
with $C_i^j$ possibly dependent on the metric. By definition, $C^i[\alpha,\beta]$ and $C_i^j$ are tensors, but is the latter concomitant of the metric? It is hard to imagine that it is not, but let us see a precise argument that it indeed has this property. We evaluate the Jacobi identity
\[
\left\{\{\alpha\cd\hh,\beta\cd\hh\},\gamma^i\cd\hh_i\right\}=\left\{\{\alpha\cd\hh,\gamma^i\cd\hh_i\},\beta\cd\hh\right\}-\left\{\{\beta\cd\hh,\gamma^i\cd\hh_i\},\alpha\cd\hh\right\},
\]
using \eqref{Cfinal1} and \eqref{Cfinal2}. We obtain after some algebra that $C^{i\j}\define C^i_kg^{jk}$ satisfies 
\eq{
\label{CJacobi}
\begin{split}
&\left((\gamma^k\partial_kC^{i\j}-C^{ik}\partial_k\gamma^j-C^{k\j}\partial_k\gamma^i)(\alpha\partial_{\j}\beta-\beta\partial_{\j}\alpha)\right)\,\cd\hh_i\\&\quad\quad\quad\quad\quad\quad=\midint_\Sigma\d^nx\,\{C^{i\j}(x),\gamma^i\cd\hh_i\}\left(\alpha(x)\partial_{\j}\beta(x)-\beta(x)\partial_{\j}\alpha(x)\right)\hh_i(x).
\end{split}
}
Note that the above equation can be written as $\int_\Sigma\xi^i\hh_i=0$, where $\xi$ is a compactly supported smooth vector field since $\alpha,\beta\!\in\!\ccc$. Integrating by parts, we get that $\int_\Sigma\nabla_{(i}\xi_{\j)}\pi^{i\j}=0$, which holds for any momentum $\pi^{i\j}$, implying that $\nabla_{(i}\xi_{\j)}=0$. For a generic metric $\xi=0$ is the only Killing vector field. Furthermore, $\alpha\partial_{\j}\beta-\beta\partial_{\j}\alpha$ can be any vector at a given point. Therefore \eqref{CJacobi} simply means that
\[
\gamma^k\partial_kC^{i\j}-C^{ik}\partial_k\gamma^j-C^{k\j}\partial_k\gamma^i=\{C^{i\j},\gamma^i\cd\hh_i\}.
\]
The left hand side is the Lie derivative of a rank two contravariant tensor, so this equation is just the mathematical expression of the fact that $C^{i\j}$ is a tensor, concomitant of the metric.

This is not the end of the story. Plugging \eqref{Cfinal2} in \eqref{HHsimpler}, and keeping only the terms that are not already of the form of an $\nn^\mu$ independent function multiplied by $\nn$, we find that
\[
\alpha C^k_{(i}\left(\partial^{}_{\j)}\nn\partial^{}_k\beta-\partial^{}_{\j)\,}\beta\partial^{}_k\nn\right)-\beta C^k_{(i}\left(\partial^{}_{\j)}\nn\partial^{}_k\alpha-\partial^{}_{\j)}\alpha\partial^{}_k\nn\right)
\]
should be proportional to $\nn$ (with proportionality factor independent of $\nn^\mu$). Let $\beta$ be constant in a neighborhood of $x$. Let us choose an $\nn$ such that $\nn(x)=0$ and all its (first order) derivatives vanish at $x$ except for $\partial_i\nn(x)$ (for a fixed $i$), and an $\alpha$ such 
that $\partial_k\alpha(x)$ is its only nonzero (first order) derivative at $x$. We find that $C_i^k=C^i_i\delta_i^k$ (no summation over $i$). Since $C^k_i$ was found to be a tensor, concomitant of the metric, this implies that
\eq{
\label{Cfinal3}
C^i[\alpha,\beta]=-\epsilon[g]g^{i\j}(\alpha\partial_{\j}\beta-\beta\partial_{\j}\alpha),
}
where $\epsilon[g]$ is a scalar, concomitant of the metric.

\vspace{1ex}   
\noindent{\it 5. Algebraic properties of $G_{i\j kl}$ and their implications}
\vspace{1ex}

\noindent Recall that we chose $\hh$ to be a tensor density of weight $1$. Since the momentum is a tensor density of weight $1$, the coefficient $G_{i\j kl}$ is a tensor with symmetry properties $G_{i\j kl}=G_{\j ikl}=G_{kli\j}$. $G_{i\j kl}$ is concomitant of the metric (this follows from the same property of $\hh$), and ultralocal in it (by assumption). A simple application of Schur's lemma in lemma \ref{lem_supermetric} shows that  
\eq{
\label{supermetric}
G_{i\j kl}=\kappa\,(g_{ik}g_{\j l}+g_{il}g_{\j k})-\lambda\,g_{i\j}g_{kl},
}
exhausts the class of such tensor densities ($\kappa$ and $\lambda$ are constant). 

Let us evaluate the Poisson brackets in \eqref{secondvar}, using \eqref{Cfinal1} and \eqref{Cfinal3}. Collecting the terms linear in the momentum, we get a quantity proportional to 
\[
\frac{\partial G_{i\j mn}}{\partial g_{kl}}\nabla_{(k}\nn_{l)}-G_{i\j k(m}\nabla_{n)}\nn^k-G_{mnk(i}\nabla_{\j)}\nn^k,%+\frac{1}{2}G_{i\j mn}\nabla_k\nn^k,
\]
which is zero since the vanishing of this quantity precisely means that the tensor $G_{i\j kl}$ is concomitant of the metric and ultralocal in it. The first term is the infinitesimal change of $G_{i\j kl}$ if we assume that it is a function of the undifferentiated metric, and the infinitesimal transformation is induced by the Lie transport of the metric along the vector field $\vec{\nn}$. The Lie derivative of a tensor with the symmetry properties of \eqref{supermetric} with respect to $\vec{\nn}$ gives the other two terms. What remains is 
\[
\begin{split}
\nn\delta_\alpha\delta_\beta g_{i\j}&=2\nn\alpha\beta\left(2\frac{\partial P_{i\j}}{\partial g_{kl}}P_{kl}-\frac{1}{\sqrt{g}}G_{i\j kl}\frac{\partial}{\partial g_{kl}}\frac{1}{\sqrt{g}}G_{mnpq}\pi^{mn}\pi^{pq}\right)-2\nn\beta\nabla_{(i}(\epsilon\nabla_{\j)}\alpha)\\
&-2\alpha\beta\left(\frac{1}{\sqrt{g}}G_{i\j kl}\frac{\delta}{\delta g_{kl}}\int_\Sigma\nn G[\g]-\nabla_{(i}(\epsilon\nabla_{\j)}\nn)\right).
\end{split}
\]
The left hand side is proportional to $\nn$, which causes the parenthesis in the second line to be a multiple of $\nn$, so
\eq{
\label{funcderG}
\frac{1}{\sqrt{g}}G_{i\j kl}\frac{\delta}{\delta g_{kl}}\int_\Sigma\nn G[\g]=\nabla_{(i}(\epsilon\nabla_{\j)}\nn)+\nn F_{i\j},
}
where the concrete form of $F_{i\j}$ is irrelevant. Recall that we required that the functional derivatives are smooth, so $G_{i\j kl}$ is contracted with a smooth tensor density on the left hand side of \eqref{funcderG}. Hence if $G_{i\j kl}(x_0)$ is not an invertible map between the spaces of rank two symmetric tensors $\mathscr{T}$ at $x_0$, then the left hand side cannot span the entire space of $\mathscr{T}$ at $x_0$ as we vary $\nn$. $G_{i\j kl}$ given by \eqref{supermetric} is invertible either everywhere ($n\lambda\neq2\kappa\neq0$) or nowhere ($n\lambda=2\kappa$ or $\kappa=0$) on $\Sigma$. In the latter case there is no place in the space where the left hand side in \eqref{funcderG} generates $\mathscr{T}$ as $\nn$ is varied. But if $\epsilon$ is not identically zero, there are points where the right hand side generates the entire $\mathscr{T}$ since the second derivatives of $\nn$ can be prescribed arbitrarily at one point. The conclusion is that either $\epsilon=0$ or $G_{i\j kl}$ is invertible.   

From now on we will consider the case when $G_{i\j kl}$ is invertible as a map in the space of rank two symmetric tensors. Let $\tilde{G}^{i\j kl}$ be the inverse of $G_{i\j kl}$. So it is not obtained by raising the indices of $G_{i\j kl}$, but by the condition $\tilde{G}^{i\j mn}G_{mnkl}=1/2(\delta^i_k\delta^j_l+\delta^i_l\delta^j_k)$, $\tilde{G}^{i\j kl}=\tilde{G}^{jikl}=\tilde{G}^{i\j lk}$. From \eqref{funcderG} we have
\eq{
\frac{1}{\sqrt{g}}\,\frac{\delta}{\delta g_{i\j}}\int_\Sigma\nn G[\g]=\tilde{G}^{i\j kl}\nabla_{(k}(\epsilon\nabla_{l)}\nn)+\nn \tilde{F}^{i\j},
\label{funcderG2}
}
where $\tilde{F}^{i\j}=\tilde{G}^{i\j kl}F_{kl}$, $\tilde{F}^{i\j}=\tilde{F}^{\j i}$. Note that \eqref{funcderG2} implies that $G[\g]$ is actually a function of the metric and its first and second derivatives (see lemma \ref{lem_field_dependence}). Using the definition of the functional derivative, we can see from \eqref{funcderG2} that for any variation $\g_\lambda$ with $\delta\g\define\partial_\lambda\g_\lambda|_{\lambda=0}$, 
\eq{
\label{funcderG3}
\frac{\d}{\d\lambda}\int_\Sigma\nn G[\g_\lambda]\bigg|_{\lambda=0}\!\!\!=\!\!\int_\Sigma\delta g_{i\j}\,\frac{\delta}{\delta g_{i\j}}\int_\Sigma\nn G[\g]. 
}
Let the variation $\g_\lambda$ be given by the diffeomorphisms generated by a vector field $\vec{\xi}$. We can exchange the order of the integral and the differentiation on the left hand side of \eqref{funcderG3}, plug in \eqref{funcderG2} on the right hand side, and integrate by parts. Considering that $\nn$ can be any compactly supported smooth function, we get that
\eq{
\label{varG}
\delta G[\g]\define\partial_\lambda G[\g_\lambda]|_{\lambda=0}=\tilde{G}^{i\j kl}\epsilon\nabla_{(k}\nabla_{l)}\delta g_{i\j}+\tilde{G}^{i\j kl}(\nabla_{(k}\epsilon)\nabla_{l)}\delta g_{i\j}+\tilde{F}^{i\j}\delta g_{i\j},
}
where we used that the covariant derivatives of $\tilde{G}^{i\j kl}$ are zero, and $\delta G[\g]$ is fixed by the condition that it is a scalar density of weight $1$, concomitant of the metric.

Now we argue that the right hand side of \eqref{varG} is inconsistent with the tensorial structure of $G[\g]$. The suspect is the term containing first order derivatives of $\delta g_{i\j}$. We have
\eq{
\label{diffG}
\partial_i(\xi^iG[\g])=2\tilde{G}^{i\j kl}\epsilon\nabla_k\nabla_l\nabla_{\j}\xi_i+2\tilde{H}^{i\j k}\nabla_k\nabla_{\j}\xi_i+2\tilde{F}^{i\j}\nabla_i\xi_{\j},
}
where we introduced the quantity $\tilde{H}^{i\j k}=\tilde{G}^{i\j kl}\nabla_l\epsilon$ with the symmetry property $\tilde{H}^{i\j k}=\tilde{H}^{\j ik}$. Since the commutators of covariant derivatives can be expressed in terms of the Riemannian tensor, \eqref{diffG} can be written as 
\eq{
\label{diffG2}
\partial_i(\xi^iG[\g])=2\tilde{G}^{i(jkl)}\epsilon\nabla_{(k}\nabla_l\nabla_{\j)}\xi_i+2\tilde{H}^{i(jk)}\nabla_{(k}\nabla_{\j)}\xi_i+K^{i\j}\nabla_i\xi_{\j}+L^i\xi_i,
}
where the concrete form of $K^{i\j}$ and $L^i$ is irrelevant. Since all the symmetric covariant derivatives of a vector field $\xi$ (including the undifferentiated $\xi$ itself) can be prescribed arbitrarily at a point (up to a fixed order), and on the left hand side of \eqref{diffG2} only $\xi$ and its first derivatives arise, \eqref{diffG2} can hold for any $\xi$ only if 
$\tilde{G}^{i(jkl)}=0$ and $\tilde{H}^{i(jk)}=0$. If $\epsilon$ is not identically zero, the first condition fixes the value of $\lambda$ in \eqref{supermetric}: $\tilde{G}^{i\j kl}=\tilde{\kappa}(g^{ik}g^{jl}+g^{il}g^{jk}-2g^{i\j}g^{kl})$ with some constant $\tilde{\kappa}$, and hence $G_{i\j kl}=\kappa\,(g_{ik}g_{jl}+g_{il}g_{jk}-\frac{2}{n-1}\,g_{i\j}g_{kl})$. Then the second condition yields $g^{jk}\nabla^i\epsilon-g^{i(j}\nabla^{k)}\epsilon=0$. By a contraction we get $(n-1)\nabla^i\epsilon=0$, so $\epsilon$ is a constant. If necessary, rescale $\hh$ by $\frac{1}{\sqrt{|\epsilon|}}$ to get $\epsilon=\pm 1$ in the Poisson algebra of $\hh_\mu$.

Finally, let us assume that $\epsilon=0$, and $G_{i\j kl}$ is still invertible. For brevity, we introduce the tensor density $t^{i\j}[\g,\alpha]=\frac{\delta}{\delta g_{i\j}}(\alpha\cd G[\g])$. The assumption $\epsilon=0$ implies that 
\eq{
\label{eo}
\midint_\Sigma\mbox{$\frac{1}{\sqrt{g}}$}\,G_{i\j kl}\,\pi^{i\j}\,(\alpha\,t^{kl}[\beta,\g]-\beta\,t^{kl}[\alpha,\g])=0.
}
The momentum $\pi^{i\j}$ can be any compactly supported symmetric tensor. The same applies to $G_{i\j kl}\pi^{i\j}$ by the invertibility of $G_{i\j kl}$. Thus \eqref{eo} implies that $t^{i\j}[\alpha,\g]=\alpha\,\tau^{i\j}[\g]$. By lemma \ref{lem_field_dependence}, $G[\g](x)$ is a function of $g_{i\j}(x)$, or in other words, $G[\g]$ is ultralocal in the metric.        
\end{proof}

Now we can decide if Ho\v{r}ava's Lagrangian, or its generalization to any (even nonlocal) potential term, is a viable modification of Einstein's gravity if we demand that such a theory should give rise to a local Hamiltonian constraint $\hh$ which -- under our conditions on the structure functions -- forms a closed Poisson algebra with the momentum constraint. In general relativity $G_{i\j kl}=\kappa\,(g_{ik}g_{\j l}+g_{il}g_{\j k}-\frac{2}{n-1}\,g_{i\j}g_{kl})$. In the class of theories we are considering the coefficient in the kinetic term is $G^{(\lambda)}_{i\j kl}=\kappa\,(g_{ik}g_{\j l}+g_{il}g_{\j k})-\lambda\,g_{i\j}g_{kl}$ with some constants $\kappa$ and $\lambda$. Since the goal is such a theory that interpolates between Einstein's gravity and a yet unknown UV theory, $\lambda$ arbitrarily approaches $\frac{2}{n-1}\,\kappa$, which means that $G^{(\lambda)}_{i\j kl}$ has to be invertible in some regime of the parameters characterizing the interpolation. The following result applies to this case: 
\begin{cor}
Let the space $\Sigma$ be $n\geq3$ dimensional, and $G_{i\j kl}$ is an ultralocal concomitant of the metric, invertible as a map between spaces of rank two symmetric tensors. Assume that the Poisson algebra of the supermomentum $\hh_i$ and
\[
\hh=\frac{1}{\sqrt{g}}G_{i\j kl}^{(\lambda)}\pi^{i\j}\pi^{kl}+G[\g]
\]
closes with the standard Poisson bracket \eqref{standard_poisson} so that the super-Hamiltonian arises in its own Poisson brackets with structure function $C[\alpha,\beta,\g,\ppi]$ which is either independent of $\pi^{i\j}$, or it depends on it at most linearly, and in this case the partial locality condition $\supp C^\mu[\alpha,\beta,\g,\ppi]\subset\supp\alpha\cup\supp\beta$ holds for all the structure functions. Then $G[\g]=\sqrt{g}\,(\mu\,R+\nu)$, where $\mu$ and $\nu$ are constant, and $R$ is the Ricci scalar of $g_{i\j}$. If $\mu\neq0$, then $G_{i\j kl}\!=\!\kappa\,(g_{ik}g_{\j l}+g_{il}g_{\j k}-\frac{2}{n-1}\,g_{i\j}g_{kl})$, where $\kappa\neq0$ is constant. 
\end{cor}
\begin{proof}
According to the theorem, if $\epsilon=0$ and $G_{i\j kl}$ is invertible, then the potential term in $\hh$ is ultralocal in the metric. So $G[\g]$ is a scalar density of weight $1$, ultralocal in the metric, hence it is $\mu\sqrt{g}$. Indeed, $\alpha^i\nabla_i\frac{1}{\sqrt{g}}G[\g]=2\Big(\frac{\partial}{\partial g_{i\j}}\frac{1}{\sqrt{g}}G[\g]\Big)\nabla_i\alpha_{\j}$ for any $\alpha^i$, which implies that $\nabla_i\frac{1}{\sqrt{g}}G[\g]=0$. If $\epsilon\neq0$, then by rescaling $\hh$ by a constant, the Poisson algebra, if it closes, can be brought into the symmetry algebra of Einstein's gravity. As shown in \refonline{teitelboim}, the only realization of this algebra with $\hh_i$ given by \eqref{eq_supmom} and an $\hh$ which is a quadratic function of the momentum is Einstein's gravity. $\epsilon\neq 0$ if and only if $\mu\neq 0$, and the theorem gives the form of $G_{i\j kl}$ in this case.
\end{proof}

\section{Conformal Lifshitz gravity}
\label{sec_uv}
The main motivation of the previous chapter was to see if there are relatively simple theories that can be considered as alternatives to Einstein's gravity. Our approach was rather conservative. The theory is a geometrodynamics, meaning that in the Hamiltonian formalism it describes the time evolution of spatial geometries, and the Hamiltonian is given by a homogeneous linear expression of local constraints whose Poisson algebra is closed. The local constraints follow from the Hamiltonian equations. We were interested in an extension of the algebra of the spatial diffeomorphisms by an additional local symmetry mostly because the possibility of a continuous deformation of the symmetry algebra of general relativity is tantalizing. In our view, Ho\v{r}ava's attempt to recover Einstein's gravity as a limit of some flow of theories can be successful only if such deformations exist. Furthermore, with the additional constraint function and the nontrivial transformations it generates on the constraint manifold, the na\"{i}ve counting of the degrees of freedom at a spacetime point yields the same number as in Einstein's gravity. This is reassuring if the degrees of freedom in Einstein's gravity are to be matched with those of the modified theory. On the other hand, as already mentioned in the Introduction and Section \ref{sec_preliminaries}, the symmetries associated with the Hamiltonian constraint make it possible in Einstein's gravity to eliminate the metric mode with negative kinetic energy. If this is considered to be the key property, one can take a less conservative standpoint. The goal can be a theory such that this mode is absent. The number of degrees of freedom per spacetime point is then the same as in general relativity, the dynamics is the time evolution of spatial geometries, but the Hamiltonian is not necessarily a combination of local constraints.    

As a starting point, we investigate the Lagrangian \eqref{UV} suggested by Ho\v{r}ava as a candidate for the UV fixed point of his theory under the anisotropic scale transformation (in $n=3$ dimensional space). The properties of this model lead us to the consideration of theories in which the only canonical variables are the conformal metric and momentum \eqref{conf_var}, and their transformations under spatial diffeomorphisms are generated by the conformal supermomentum \eqref{eq_supmom_conf}. 

In \eqref{UV} the kinematically possible configurations of $\nn$ cannot be restricted to spatial constants since the scale transformation of $\nn$ proposed by Ho\v{r}ava is space dependent. The variation of \eqref{UV} with respect to $\nn$ gives a local constraint which does not form a closed Poisson algebra with the conformal momentum constraint. The simplest way of getting a closed Poisson algebra of constraints is to demand that $\nn$ is a (spatial) constant, so the only local constraint is the conformal supermomentum. We show that this restriction is not an obstacle to the realization of the local Weyl symmetry. We will modify \eqref{UV} so that the action in the Hamiltonian formalism depends only on the conformal variables. Local scale transformations act on these variables trivially, so they leave the action invariant.

The action \eqref{UV} can be got from the following action if we eliminate $\pi^{i\j}$, using the field equation obtained by the variation with respect to $\pi^{i\j}$, with the assumption that no spatial boundary term arises:
\eq{
\label{uv_action_ham}
S_\mathrm{UV}=\!\!\mathop{\mbox{$\int$}}_{\mathbb{R}\times\Sigma}\left(\tilde{\pi}^{i\j}\dot{\tilde{g}}_{i\j}-\nn^i\tilde{\hh}_i-\nn\hh_\mathrm{UV}\right),
}
where we used the conformal variables \eqref{conf_var}. For simplicity we set $\kappa^2=2$. $\tilde{\hh}_i$ is the conformal supermomentum \eqref{eq_supmom_conf}, and
\eq{
\label{uv_ham}
\hh_\mathrm{UV}=\frac{1}{\sqrt{g}}\bigg(\pi^{i\j}\pi_{i\j}-\frac{1}{3}\pi^2\bigg)+\frac{\sqrt{g}}{w^4}C^{i\j}C_{i\j}.
}
The first term in \eqref{uv_action_ham} is a nondegenerate bilinear expression of $\tilde{g}_{i\j}$ and $\tilde{\pi}^{i\j}$, so the canonical variables are the conformal metric and momentum. This term has the standard form $\smallint_\Sigma\pi^{i\j}\dot{g}_{i\j}$ for which the Poisson brackets are defined by \eqref{standard_poisson}, but here $(g_{i\j},\pi^{i\j})$ are subject to the second class constraints $g\!=\!1$ and $\pi\!=\!0$, so the bracket is the corresponding Dirac bracket. Since $g$ and $\pi$ have vanishing Poisson brackets with $(\tilde{g}_{i\j},\tilde{\pi}^{i\j})$, the Dirac bracket is the standard Poisson bracket written in terms of the conformal variables and $(g,\pi)$, without the term containing functional derivatives with respect to the latter variables.          
 
From the perspective of Dirac's approach to constrained systems (see for example \refonline{HRT}) it would be more natural to include $\pi$, the generator of the local scale transformations of the spatial metric, in the Hamiltonian as one of the constraints:
\eq{
\label{uv_action_mod}
S^\prime_\mathrm{UV}=\!\!\midint_{\mathbb{R}\times\Sigma}\left(\pi^{i\j}\dot{g}_{i\j}-\nn^i\hh_i-\lambda\,\pi\right)-\midint_\mathbb{R}H,
}
where $\lambda$ is a scalar function. Since $\pi$ is spatial scalar density, the Poisson algebra of $(\pi,\hh_i)$ closes. We also included a term $H$. If the Poisson brackets of $H$ with the constraint functions $\hh_i$ and $\pi$ are their homogeneous combinations, there is no need for imposing further local constraints in order to guarantee that the time evolution determined by \eqref{uv_action_mod} preserves $\pi=0$ and $\hh_i=0$. Tentatively, we can write $H=\nn\cd\hh_\mathrm{UV}$. All the symmetry generators are already present in the Hamiltonian of \eqref{uv_action_mod}, so we do not lose anything, but gain a closing Poisson algebra of the constraints $(\pi,\hh_i)$ if we modify the action \eqref{uv_action_mod} so that $\hh_\mathrm{UV}$ is not a local constraint. First we replace the space dependent $\nn$ by a constant $\nu$, otherwise the variation of \eqref{uv_action_mod} with respect to $\nn$ would give $\hh_\mathrm{UV}=0$. But this is not enough in order to avoid $\hh_\mathrm{UV}=0$ as the consequence of the field equations. Using the scaling properties of the Cotton tensor (see below equation \eqref{UV}) algebra, we can see that
\eq{
\label{scaling_hh}
\left\{\alpha\cd\hh_\mathrm{UV},\pi(x)\right\}=-\frac{3}{2}\alpha(x)\hh_\mathrm{UV}(x),
}
which means that $\pi=0$ forces us to impose the Hamiltonian constraint $\hh_\mathrm{UV}=0$ as well, otherwise $\pi$ would not be preserved by the time evolution. (A negative constant can be added to $\hh_\mathrm{UV}$ in order to avoid the constraint $\pi^{i\j}=0$.) We can exclude the possibility that the Poisson bracket of $\alpha\cd\hh_\mathrm{UV}$ and $\beta\cd\hh_\mathrm{UV}$ closes on the supermomentum. Indeed, if that was the case, then by theorem \ref{main}, this Poisson bracket should be zero for any $\alpha$ and $\beta$ since $G_{i\j kl}$ is not invertible. On the other hand, \eqref{scaling_hh} implies that $({\hh'}_{\!\!\!\!\!\mathrm{UV}}\!\define\!\hh_\mathrm{UV}+\frac{1}{\sqrt{g}}\pi^2,\hh_i)$ would form a Poisson algebra with the same structure functions as those of the algebra of $(\hh_\mathrm{UV},\hh_i)$. But the coefficient in the kinetic term of ${\hh'}_{\!\!\!\!\!\mathrm{UV}}$ is invertible, so by the theorem again, the potential term in ${\hh'}_{\!\!\!\!\!\mathrm{UV}}$ should be an ultralocal functional of $g_{i\j}$, and $C^{i\j}C_{i\j}$ is not of this form. This argument applies to any Weyl covariant nontrivial potential. 

If our goal is such a theory in which the temporal diffeomorphism symmetry is replaced by a local scale transformation of the spatial metric, then there is a simple way to save \eqref{uv_action_ham}. Since $\hh_\mathrm{UV}=0$ is the field equation that we get by varying \eqref{uv_action_mod} with respect to the local scale factor of the spatial metric, we will eliminate this degree of freedom by imposing the constraint $\ln g=0$. The action is
\eq{
\label{uv_action_mod2}
\tilde{S}_\mathrm{UV}=\!\!\mathop{\mbox{$\int$}}_{\mathbb{R}\times\Sigma}\left(\pi^{i\j}\dot{g}_{i\j}-\nn^i\hh_i-\lambda\,\pi-\gamma\ln g\right)-\mathop{\mbox{$\int$}}_{\mathbb{R}\times\Sigma}\!\nu\,\hh_\mathrm{UV},
}
where $\nu$ is a (spatial) constant. The field equations of \eqref{uv_action_mod2} after the elimination of $\lambda$ and $\gamma$ are equivalent to the Hamiltonian equations of
\eq{
\label{uv_ham_mod}
\tilde{H}_\mathrm{UV}=-2\int_\Sigma\nn^i\tilde{\nabla}_{\j}\tilde{\pi}^j_i+\nu\int_\Sigma\bigg(\tilde{\pi}^{i\j}\tilde{\pi}_{i\j}+\frac{1}{w^2}\tilde{C}^{i\j}\tilde{C}_{i\j}\bigg)
}
with the standard brackets \eqref{standard_poisson} replaced by the Dirac brackets for the constraints $\pi=0$, $\ln g=0$. $\tilde{\nabla}$ is the covariant derivative compatible with $\tilde{g}_{i\j}$, $\tilde{C}^{i\j}$ is the Cotton tensor of $\tilde{g}_{i\j}$, and the indices are lowered by $\tilde{g}_{i\j}$. The scale transformation acts trivially on the conformal variables \eqref{conf_var}. Finally we note that na\"{i}ve counting of the degrees of freedom of \eqref{uv_ham_mod} at a spacetime point gives the same result as in Einstein's gravity. The number of the canonical pairs is less by one, but the same applies to the number of constraint functions and gauge fixing conditions. 
\section{Discussion}
\label{sec_disc}
In Einstein's gravity, the ultralocal metric dependence of the kinetic term in $\hh$ is the consequence of the Poisson algebra of $\hh_\mu$.\cite{teitelboim} Allowing the kinetic term to depend on the derivatives of the metric might open the way to new extensions of the spatial diffeomorphisms. Probably, the relaxation of time reversibility is the most interesting generalization. This would mean a term in $\hh$ linear in the momentum.\cite{HKT} As we mentioned earlier, there is no irreversible Einsteinian geometrodynamics.\cite{kuchar} It is an intriguing question if irreversibility allows for a symmetry algebra other than that of general relativity or ultralocal gravity.

\section*{Acknowledgment}
We wish to thank Patrick Draper for useful comments on the manuscript. This work was supported in part by DOE grant DE-FG02-90ER-40560.

\setcounter{section}{0}
\titlelabel{Appendix \thetitle:\hspace{6pt}}
\renewcommand\thesection{\Alph{section}}
\section{Notations, assumptions, and their immediate consequences}
\label{assumptions_app} 
The space $\Sigma$ is an $n$ dimensional smooth Riemannian manifold with a metric $\g$ on it. It is not required to be compact, and it can have a boundary. Let $\Omega\subset\Sigma$ be a bounded domain (i.e.\ a connected open set whose closure is compact). The boundary $\partial\Omega$ of $\Omega$ is always assumed to be a smooth manifold. $C^\infty(\cO)$ is the space of smooth functions on $\cO$, and its subspace $\cc{\O}$ consists of functions whose support is within $\Omega$. The space of smooth vector, $m$-index, or symmetric two-index tensor fields is denoted by $C^\infty(\cO,T)$, $C^\infty(\cO,\otimes_mT)$, or $C^\infty(\cO,T\!\vee\!T)$. We introduce the spaces $\cc{\O,T}$, \ldots similarly to the scalar functions. $C^\infty(\Sigma)$ is the space of the smooth functions on $\Sigma$, and $\cc{\Sigma}$ is the subspace of functions whose support is in the interior of $\Sigma$. 

The $p$-integrable functions, vector fields, \ldots live in $L^p(\Omega)$, $L^p(\Omega,T)$, \ldots\@ For $p=2$ they are Hilbert spaces over $\mathbb{R}$. $H^\ell(\Omega)$, $H^\ell(\Omega,T)$, \ldots are the Sobolev spaces, and $\sob{\ell}(\Omega)$, $\sob{\ell}(\Omega,T)$, \ldots are the completion of $\cc{\O}$, $\cc{\O,T}$, \ldots in them. These are also Hilbert spaces, but we will use only the norm, and never the scalar product on them. Thus $\langle,\rangle$ always refers to the $L^2$ scalar product, defined as $\langle\mathbold{s},\mathbold{t}\rangle\define\smallint_\Omega s^{i_1\dots i_m}t_{i_1\dots i_m}$ for $\mathbold{t},\mathbold{s}\in L^2(\Omega,\otimes_mT)$ even if the arguments happen to be in $H^\ell\subset L^2$. Recall the definition of the Sobolev spaces $H^\ell(\Omega)$. (The other spaces $H^\ell(\Omega,T)$, \ldots are defined in the same way.) $H^0(\Omega)=L^2(\Omega)$, and the elements of $H^\ell(\Omega)$ ($\ell\geq 1$) are the functions in $H^{\ell-1}(\Omega)$ whose weak derivative of order $\ell$ exists, and it is square integrable, i.e., it is in $L^2(\Omega,\otimes_\ell T)$. The standard norm on $H^\ell(\Omega)$ is defined by $\|f\|_{H^\ell(\Omega)}^2\define\|f\|_{H^{\ell-1}(\Omega)}^2+\|\underbrace{\vec{\nabla}\dots\vec{\nabla}}_\ell f\|^2_{L^2(\Omega,\otimes_\ell T)}$.

We will also use the Banach spaces $C^r(\cO)$ of $r$ times continuously differentiable functions. For $r=0$ this is the space of the continuous functions, which is usually denoted by $C(\cO)$. As usual, the symbol $C^r(\cO,T)$, \ldots stands for the vector and tensor fields. The norms on these spaces are $\|f\|_{C^r(\cO)}=\max\limits_{0\le s\le r}\max\limits_{x\in\cO}|\underbrace{\vec{\nabla}\dots\vec{\nabla}}_{s}f(x)|$, etc., where $|\mathbold{t}(x)|$ can be $|\mathbold{t}(x)|^2=t_{i_1\dots i_m}(x)t^{i_1\dots i_m}(x)$. 

The valence of the elements of all these spaces must be fixed, otherwise addition on them would not make any sense, but a field whose valence is different from our convention can be considered as an element of our space by raising or lowering the appropriate indices. We can define a tensor field $\pp$ by $\sqrt{g}\,\pp=\ppi$. We will use $\pp$ instead of the momentum $\ppi$ so that we have to deal with only tensors, and not with tensor densities as well.

A kinematically allowed configuration of the canonical variables is a configuration for which the (closed) Poisson algebra generated by $\{\,\alpha^\mu\cdot\hh_\mu\,|\,\alpha^\mu\in\cc{\Sigma}\,\}$ is defined. ($\alpha^\mu$ are independent of the canonical variables.) Apart from smoothness requirements, these configurations are subject to some boundary conditions, or they have specific decaying properties if $\Sigma$ has a boundary, or it is not compact. We do not need a detailed definition of their manifold, but we assume that locally the metric and the momentum can be anything, and the manifold of the metric configurations has a kind of topological property that makes it possible to infer local metric dependence from the functional derivatives of certain functionals:
\renewcommand{\theenumi}{\roman{enumi}}
\begin{enumerate}
\item\label{kinematics_prop} Let $\G=\mathcal{G}\times\mathcal{P}$ be the manifold of the kinematically allowed configurations of the canonical variables, where $\mathcal{G}$ is the space of the kinematically allowed metrics, $\mathcal{P}$ is that of the momenta. $\cc{\S,T\!\vee\!T}\subset\mathcal{P}$ is a vector space. Let $\g_1$ and $\g_2$ be two metrics on $\Sigma$, $\g_1\in\mathcal{G}$, and $\g_1-\g_2\in\cc{\S}$. Then we assume that $\g_2\in\mathcal{G}$. Furthermore, any $\g_1,\g_2\in\G$ are connected by a variation $(\f_\lambda)_{\lambda\in(0,1)}\in\G$ which is constant on $\Sigma\setminus\supp(\g_1-\g_2)$. 
\end{enumerate}
When we say momentum, we always mean a smooth tensor field $\pp$ which is kinematically allowed, $\pp\in\mathcal{P}$. Sometimes the word ``momentum'' refers to the corresponding tensor density $\ppi=\sqrt{g}\,\pp$. (The momentum conjugate to the metric is actually a tensor density.) The momentum $\pp$ is said to be transverse (on $\Omega$) if $\nabla_ip^{i\j}=0$ (on $\Omega$). 

We will impose some continuity conditions on the quantities arising in the Poisson algebra. Let $\O\subset\S$ be open. $\mathscr{D}(\O)$ is the vector space of the compactly supported smooth functions on $\O$ with the topology in which a sequence $f_k$ goes to zero if there is a compact set $K\subset\O$ such that $\supp f_k\subset K$ for all $k$, and $f_k$ and all its derivatives uniformly tend to zero. A sequence $f_k$ is said to converge to $f$ in $\mathscr{D}(\O)$, if $f_k-f$ goes to zero in $\mathscr{D}(\O)$. The spaces $\mathscr{D}(\O,T)$, \ldots and the convergences in them are defined similarly. The conditions:
\begin{enumerate}
\setcounter{enumi}{1}
\item\label{cont_structfunc_prop} Let $\O\subset\S$ be open. If $(\alpha,\pp)\mapsto C^\mu[\alpha,\beta,\g,\pp]|_\O$ is a map $\cc{\O}\times\cc{\S,T\!\vee\!T}\to\cc{\O}$ for all $\g\in\mathcal{G}$ and $\beta\in\cc{\S}$, then it is required to be continuous in the $\mathscr{D}$-topology.
\item\label{cont_functional_prop} %Let $F:\mathcal{G}\times\mathcal{P}\to\mathbb{R}$ be an element of the Poisson algebra, $\g_k$ a sequence in $\mathcal{G}$ such that $\g_k$ and $\g^{-1}_{k}$ both converge in $\mathscr{D}(\S)$. Then $\lim_{k\to\infty}F[\g_k,\pp]=F[\lim_{k\to\infty}\g_k,\pp]$ for any $\pp\in\mathcal{P}$.
Let $F:\mathcal{G}\times\mathcal{P}\to\mathbb{R}$ be an element of the Poisson algebra. Then $\g\mapsto F[\g,\pp]$ is required to be continuous in the $\mathscr{D}$-topology for any $\pp\in\mathcal{P}$.
\item\label{anal_prop} If $\g$ is analytic in local coordinates $(\O,\varphi)$, so is the potential function $G[\g]$ of $\hh$.
\end{enumerate}
The space $\Sigma$ is defined to be a smooth, but not an analytic manifold. That is why analiticity of functions and tensors is a coordinate dependent property, as in the last condition. These assumptions are not restrictive at all for practical purposes. If $G[\g](x)$ is a polynomial of the metric, its inverse, and its derivatives at $x\in\Sigma$ up to an arbitrarily high fixed order with coefficients independent of $x$, then $\alpha\cd G[\g]$ satisfies conditions \eqref{cont_functional_prop} and \eqref{anal_prop} for any $\alpha\in\cc{\Sigma}$ that does not depend on the canonical variables. For theorem \ref{main} we need only the first two properties. The other two are used only for theorem \ref{linC}. Condition \eqref{cont_functional_prop} makes it possible to extend certain results obtained for locally analytic metrics to any kinematically allowed metric since we have
\begin{lem}
\label{analytic_approx_lem}
Let $\Omega_0$ be a bounded domain for which there is a coordinate chart $(\Omega,\varphi)$ so that $\cO_0\subset\Omega$. For any $f\in C^\infty(\Sigma)$ there is a sequence $f_k$ which is analytic on $\Omega_0$ in the given local coordinates and tends to $f$ in $\mathscr{D}(\Sigma)$ as $k\to\infty$.
\end{lem}
\begin{proof}
We can assume that $\Omega$ is bounded. There is a compactly supported smooth $F:\mathbb{R}^n\to\mathbb{R}$ which is equal to $f\circ\varphi^{-1}$ on $\varphi(\Omega)$. The norm defined by $\|F\|^2=\smallint_{\mathbb{R}^n}\d^nx(1+|x|^2)^\ell|\hat{F}(x)|^2$ is equivalent to the standard norm $\|\cdot\|_{H^\ell(\mathbb{R}^n)}$, where $\hat{F}$ is the Fourier transform of $F$ on the Euclidean space $\mathbb{R}^n$. Since $\|F\|_{H^\ell(\mathbb{R}^n)}<\infty$ for any $\ell\geq0$, $\lim_{k\to\infty}\smallint_{|x|>k}\d^nx(1+|x|^2)^\ell|\hat{F}(x)|^2=0$. Let $\hat{F}_k=\hat{F}(x)$ if $|x|<k$, and $\hat{F}_k(x)=0$ otherwise. $F_k$, the inverse Fourier transform of $\hat{F}_k$, is real analytic for any $k$. There is a smooth function $\rho:\Omega\to\mathbb{R}$ which is $1$ on $\Omega_0$ and $\supp\rho\subset\Omega$. Define $f_k$ on $\Omega$ by $f_k=\rho\,(F_k\circ\varphi)+(1-\rho)f$, and let $f_k=f$ otherwise. By the Sobolev embedding theorem, if $2(\ell-r)>n$, then $H^\ell(\Omega)\hookrightarrow C^r(\cO)$, and this embedding is continuous. Thus the sequence $f_k$ has the desired property since all its derivatives uniformly converge.
\end{proof}
The lemma we just proved naturally extends to vector and tensor fields, including nondegenerate tensor fields like metrics. One example why \eqref{cont_structfunc_prop} can be useful:
\begin{lem}
\label{lem_local}
Let $\O\!\subset\!\S$ be open, $H\!:\!\mathscr{D}(\O)\!\to\!\mathscr{D}(\O)$ a continuous linear map, and $\supp H[\alpha]\!\subset\!\supp\alpha$ for any $\alpha\in\cc{\O}$. Then at any point $x\in\O$ there are smooth functions $C^{J,i_1i_2\dots i_m}$ so that $H[\alpha](x)$ and its derivatives $\partial_\mathbf{j}H[\alpha](x)$ (where $\mathbf{j}$ is a multiindex) can be written for all $|\mathbf{j}|\!\le\! J\!<\!\infty$ as
\[
\partial_\mathbf{j}H[\alpha](x)=\sum_{m=0}^{M^J(x)}\partial_\mathbf{j}(C^{J,i_1i_2\dots i_m}\partial_{i_1}\partial_{i_2}\dots\partial_{i_m}\alpha)(x),
%\;\;\;|\mathbf{j}|\le J,
\]
where $M^J(x)<\infty$ for all $x$ and $J$.
\end{lem}
\begin{proof}
Define
\[
\langle T_x,\alpha\rangle\define H[\alpha](x),\;\;\;\;\alpha\in \cc{\O}.
\]
By the assumption on $f$, $T_x$ is a distribution. Furthermore, if $x\notin\supp\alpha$ then $x\notin\supp H[\alpha]$, so $\langle T_x,\alpha\rangle$=0. Thus $\supp T_x\subset\{x\}$. Any distribution whose support is (at most) a point is a (finite) combination of the Dirac-delta and its derivatives at that point:
\eq{
\label{txexpand}
\begin{split}
\langle T_x,\alpha\rangle&=\sum_{m=0}^{M^0(x)}(-1)^mC^{i_1i_2\dots i_m}(x)\langle\partial_{i_1}\partial_{i_2}\dots\partial_{i_m}\delta_x,\alpha\rangle\\&=
\sum_{m=0}^{M^0(x)}C^{i_1i_2\dots i_m}(x)\,\partial_{i_1}\partial_{i_2}\dots\partial_{i_m}\alpha(x),\;\;\;\;M^0(x)<\infty,\;\;\;\alpha\in \cc{\O}.
\end{split}
} 
Note that we have not proved that $M^0(x)$ is bounded in some neighborhood of any $x$, so we have not shown that any derivative of $H[\alpha]$ can be got by differentiating the formula for $\mathbf{j}=0$. But we do not need this. As a $\mathscr{D}(\O)\to\mathscr{D}(\O)$ map, any derivative of $H[\alpha]$ has the same properties as $H[\alpha]$ itself, so it can be written as \eqref{txexpand}: $\partial_\mathbf{j}H[\alpha](x)=\sum_{m=0}^{M^\mathbf{j}(x)}{C_\mathbf{j}}^{\!\!i_1i_2\dots i_m\,}(x)\,\partial_{i_1}\partial_{i_2}\dots\partial_{i_m}\alpha(x)$, where $M^\mathbf{j}(x)<\infty$. $\partial_\mathbf{j}H[\alpha](x)$ contains only a finite number of derivatives of $\alpha$. In a neighborhood of $x$ let $\tilde{\alpha}\in\cc{\O}$ be the Taylor series of $\alpha$ about $x$ truncated at $M^J(x)$, which is the order of the highest derivative of $\alpha$ arising in $\partial_\mathbf{j}H[\alpha](x)$, $|\mathbf{j}|\le J$. Clearly, $\partial_\mathbf{j}H[\alpha](x)=\partial_\mathbf{j}H[\tilde{\alpha}](x)$ if $|\mathbf{j}|\le J$. But the sum extends over $m\le M^J(x)$ in a neighborhood of $x$ if it is evaluated on $\tilde{\alpha}$. The coefficients $C^{i_1\dots i_m}$ are smooth since $H[\alpha]$ is smooth for any $\alpha$, and $\partial_\mathbf{j}H[\tilde{\alpha}](x)$ is simply the derivative of the formula for $H[\tilde{\alpha}]$, which is a finite combination of the derivatives of $\tilde{\alpha}$ with smooth coefficients in a neighborhood of $x$. 
\end{proof}
\section{No hidden constraints from spatial diffeomorphisms}
\label{supmom_app}
Let $S_{\!\g}$ be a subspace of compactly supported smooth symmetric tensor fields on the space $\Sigma$. Subscript $\g$ indicates that the $S_{\!\g}$ might depend on the metric. Suppose that $S_{\!\g}$ is orthogonal to all the transverse momenta. In other words, the constraints $\smallint_\Sigma t_{i\j}p^{i\j}$ ($\mathbold{t}\in S_{\!\g}$) are consequences of the momentum constraint: $\hh_i=0$ implies that $\smallint_\Sigma t_{i\j}p^{i\j}=0$ for all $\mathbold{t}\in S_{\!\g}$. (The tensor $\pp$ is related to the canonical momentum by $\ppi=\sqrt{g}\,\pp$.) Is it possible to express the latter constraints in terms of the supermomentum? Is there a smooth vector field $\vec{v}$ for any $\mathbold{t}\in S_{\!\g}$ such that $\smallint_\Sigma t_{i\j}p^{i\j}=\smallint_\Sigma v_i\nabla_{\j}p^{i\j}$, and the support of $\vec{v}$ is within that of $\mathbold{t}$? Let the formal adjoint of the differential operator $\nabla_ip^{i\j}$ be $\K$, which is proportional to the Killing operator defined on compactly supported vector fields. It would be enough to show that $S_{\!\g}\subset\ran\K$. This is not obvious at all as the  relationship $\ker\K^+=(\ran\K)^\perp$ might suggest. One problem is that we should define the Hilbert space in which the orthogonal complement, the adjoint, and later the closure are taken. Furthermore, $\K^+$ is not a formal adjoint, but the actual one, and it is likely to be defined on a space bigger than that of the smooth vector fields, so it might be an extension of $\nabla_ip^{i\j}$. Thus $(\ker\K^+)^\perp$ might be smaller than $S_{\!\g}$. Finally, what we have is $(\ker\K^+)^\perp=\overline{\ran \K}$, and it is not obvious why the smooth tensor fields in the closure $\overline{\ran\K}$ should be in fact in $\ran\K$. These are the details that are worked out in the lemma in this appendix. But before the lemma, we need to ask our question in a more precise way. 

Later it will suffice if certain constraints that follow from the momentum constraint can be expressed in terms of the supermomentum only locally, so first we reformulate our question accordingly. Let $\Omega\subset\Sigma$ be a bounded domain. Define the linear functional $L:C^\infty(\cO,T\!\vee\!T)\to\mathbb{R}$ by $L[\pp]\define\smallint_\Omega t_{i\j}p^{i\j}$, where $\mathbold{t}\in \cc{\O,T\!\vee\!T}$. Let $\D:C^\infty(\cO,T\!\vee\!T)\to C^\infty(\cO,T)$ be the supermomentum considered as a linear functional of $\pp$, that is, $D^j[\pp]=\nabla_i p^{i\j}$. If $\ker\D\subset\ker L$, is there a $\vec{v}\in\cc{\O,T}$ such that $L[\pp]=\smallint_\Omega v_i\nabla_{\j}p^{i\j}$? This is the relevant question in the context of the paper, but as it frequently happens, it is easier to give an affirmative answer to it if we rephrase it on more convenient spaces, which are larger than those of the smooth functions. We will impose on $L$ a kind of continuity condition with respect to the operator $\D$. The expression $L[\pp]\define\smallint_\Omega t_{i\j}p^{i\j}$ defines a continuous functional on $L^2(\Omega,T\!\vee\!T)$. The operator $\D$, considered as a map from $L^2(\Omega,T\!\vee\!T)$ into $L^2(\Omega,T)$, is defined on a dense subspace, and its closure $\cD$ exists since the domain of $\D^+$ is also dense in $L^2(\Omega,T)$. For example, $\mathop{\mathrm{dom}}\D^+$ surely contains $\cc{\O,T}$, on which $\D^+$ acts as $\vec{\xi}\mapsto-1/2\,\mathcal{L}_{\vec{\xi}}\g$. Does $\ker\cD\subset\ker L$ imply that $L[\pp]=\smallint_\Sigma v_i\nabla_{\j}p^{i\j}$ with some $\vec{v}\in \cc{\O,T}$?

The additional continuity property of $L$ means that we not only require that $L$ is zero whenever $\D$ is zero, but $L$ also vanishes on tensors which are the limits of sequences of smooth tensor fields along which $\D$ goes to zero. Note that it is important that $\vec{v}$ whose existence our question addresses is required to be supported within $\supp\mathbold{t}$. The following statement will suffice for our purposes.

\begin{lem}
\label{supmom_unique_lem}
Let $\Omega$ be a sufficiently small bounded star domain with smooth boundary $\partial\Omega$. The map $L:L^2(\Omega,T\!\vee\!T)\to\mathbb{R}$ is defined by $L[\pp]=\smallint_\Omega t_{i\j}p^{i\j}$, where $\mathbold{t}\in \cc{\O,T\!\vee\!T}$, and the operator $\D:L^2(\Omega,T\!\vee\!T)\rightarrowtail L^2(\Omega,T)$ is given by $\mathop{\mathrm{dom}}\D=C^\infty(\cO,T\!\vee\!T)$, $D^j[\pp]=\nabla_ip^{i\j}$. If $\ker\cD\subset\ker L$, then there exists a $\vec{v}\in \cc{\Omega,T}$ such that $L[\pp]=\smallint_\Omega v_i\nabla_{\j}p^{i\j}$ for any $\pp\in C^\infty(\cO,T\!\vee\!T)$.  
\end{lem}
\begin{proof}
Let $\K\!:\!L^2(\Omega,T)\!\rightarrowtail\!L^2(\Omega,T\!\vee\!T)$, $\mathop{\mathrm{dom}}\K=\cc{\Omega,T}$, $\K[\vec{\xi}^{}]\!=\!-1/2\,\mathcal{L}_{\vec{\xi}}\g$. We have $\K^+\supset\cD$ since $\K^+$ is closed, and $\K^+|_{{C_{}}^\infty(\cO,T\vee T)}=\D$. ($A\supset B$ means that $\dom A\supset\dom B$ and $A|_{\dom B}=B$.) Actually, $\K^+=\cD$. The argument is borrowed from the theory of Sobolev spaces. The existence of the derivative $w^j=\nabla_ip^{i\j}$ in the weak sense supplemented with the square integrability of $\pp$ and $\vec{w}$ is equivalent to $\pp\in\mathop{\mathrm{dom}}\K^+$. Along the lines of the argument that shows that $C^\infty(\cO)$ is dense in the Sobolev spaces $H^\ell(\Omega)$ (this is where the star-convexity of $\Omega$ comes into play), here we get that the elements of $\mathop{\mathrm{dom}}\K^+$ are the limits of convergent sequences $\pp_k\in C^\infty(\cO,T\!\vee\!T)$ for which $\nabla_ip_k^{i\j}$ is also convergent. This precisely means that $\mathop{\mathrm{dom}}\K^+\subset\mathop{\mathrm{dom}}\cD$.    

Since $\ker\K^+=(\ran\K)^\perp$, where $\perp$ denotes the orthogonal complement, $(\ker\K^+)^\perp=\overline{\ran\K}$, and therefore $(\ker\cD)^\perp=\overline{\ran\K}$. By the assumption on $L$, the tensor field $\mathbold{t}$ is orthogonal to $\ker\cD$, hence $\mathbold{t}\in\overline{\ran\K}$. What we have to prove is that in fact $\mathbold{t}\in\ran\K$, or in other words, that $\overline{\ran\K}\cap\cc{\Omega,T\!\vee\!T}=\ran\K$.

Let $\vec{\xi}_k\in\cc{\O,T}$ be a sequence such that $\K[\vec{\xi}_k]$ is convergent in $L^2(\Omega,T\!\vee\!T)$. First we want to deduce the convergence of $\vec{\xi}_k$ from that of $\mathcal{L}_{\vec{\xi}_k}\g$. The Riemannian version of Korn's inequality\cite{chen_jost} states that there is a constant $C$ such that 
\eq{
\label{korn}
\|\vec{\nabla}\vec{\xi}^{}\|^2\le C(\|\vec{\xi}^{}\|^2+\|\mathcal{L}_{\vec{\xi}}\g\|^2).
}
for any $\vec{\xi}\in C^\infty(\cO,T)$ (and hence for any $\vec{\xi}\in H^1(\Omega,T)$). (The condition on $\partial\Omega$ can be weakened. See \refonline{chen_jost} for the details.) The symbol $\vec{\nabla}\vec{\xi}$ denotes the rank two tensor of the covariant derivative of $\vec{\xi}$. By Friedrichs' inequality,
\eq{
\label{friedrichs0}
\|\xi\|_0\le(\mathop{\mathrm{diam}}\nolimits_0\Omega)\|\vec{\nabla}_{\!0\,}\vec{\xi}\|_0
}
for any for any $\vec{\xi}\in\cc{\O,T}$ (and hence for any $\vec{\xi}\in\sob{1}(\Omega,T)$). The subscripts $0$ in \eqref{friedrichs0} indicate that the norms and the derivatives are calculated on the flat Euclidean background. Note the boundary condition on $\xi$  for Friedrichs' inequality. The coefficient $\mathop{\mathrm{diam}_0}\Omega$ is the diameter of $\Omega$, that is, the longest distance between two points of the boundary. The boundedness of the metric and its derivatives on $\cO$ allows for the following straightforward, even if perhaps not the most efficient generalization of Friedrichs' inequality to a Riemannian manifold:
\eq{
\|\xi\|^2\le c(\mathop{\mathrm{diam}}\Omega)^2(\|\vec{\nabla}\vec{\xi}\|^2+\|\vec{\xi}\|^2),
\label{friedrichs}
}
where $c$ is a constant. This and the other constant $C$ in \eqref{korn} can depend on $\g$ and $\Omega$. The derivation of \eqref{friedrichs} from \eqref{friedrichs0} also shows that $c$ can be chosen so that it depends on the metric $\g$ only through a continuous function of the maximum of the Christoffel symbols on $\cO$ ($\max_{\cO}\Gamma^i_{jk}g^{jm}g^{kn}\Gamma^l_{mn}$), $\max_{x\in\cO}\g(x)$, and $1/\min_{x\in\cO}\g(x)$, where the extrema of the matrices in a given coordinate system are the minimum and the maximum of their eigenvalues on $\cO$. Let $\Omega_0$ be a bounded domain with smooth boundary, and assume that all the domains $\Omega$ we are considering here are within $\Omega_0$. If $\vec{\xi}\in\cc{\Omega}$, then the extension $\vec{\xi}_0$ is in $\cc{\O_0}$, where $\vec{\xi}_0|_\Omega=\vec{\xi}$, and $\vec{\xi}_0|_{\Omega_0\setminus\Omega}=0$. This means that similarly to $c$, the constant $C$ can also be chosen to be $\Omega$-independent so that \eqref{korn} and \eqref{friedrichs} hold for any $\Omega\subset\Omega_0$. Thus the combination of these inequalities results in an upper bound on the Sobolev norm $\|\vec{\xi}\|_{H^1(\Omega,T)}$ of $\vec{\xi}\in\cc{\Omega,T}$, provided that $\Omega$ is sufficiently small:
\[
%\label{sobolev}
\|\vec{\xi}\|_{H^1(\Omega,T)}^2\define\|\vec{\xi}\|^2+\|\vec{\nabla}\vec{\xi}\|^2\le\gamma\|\mathcal{L}_{\vec{\xi}}\g\|^2,
\]
where the constant $\gamma$ depends on the metric and the domain. Since $\mathcal{L}_{\vec{\xi}_k}\g$ converges in $L^2(\Omega,T\!\vee\!T)$, so does $\vec{\xi}_k$ in $H^1(\Omega,T)$.
 
Next, we will show that $\vec{\xi}\define\lim_{k\to\infty}\vec{\xi}_k\in C^\infty(\Omega,T)$. Since the limit is taken in $H^1(\Omega,T)$, $\vec{\xi}$ is also in $H^1(\Omega,T)$. In particular, the first weak derivatives of $\vec{\xi}$ exist, they are the limits of the classical derivatives of $\vec{\xi}_k$ in $L^2(\Omega,T\otimes T)$, so we have $\nabla_{\!i\,}\xi_{\j}+\nabla_{\!\j\,}\xi_i=-t_{i\j}$, where the derivatives of $\vec{\xi}$ on the left hand side are defined weakly. Thus
\eq{
\langle\nabla_is^{i\j},\xi_{\j}\rangle=\langle s^{i\j},t_{i\j}\rangle-\langle\nabla_is^{ji},\xi_j\rangle\;\;\;\;\mbox{for any}\;\;\;\mathbold{s}\in\cc{\O,T\otimes T}.
\label{killing_weak}
}
The smoothness of $\vec{\xi}$ can be shown by induction. Let ${R_{i\j k}}^l$ be the Riemannian curvature tensor. Following the steps by which $\nabla_i\nabla_{\j}\xi_k=-{R_{\j ki}}^l\xi_l$ is derived for a Killing vector field, we can prove that the second weak derivative of $\vec{\xi}$ exists. We start with
\[
\langle(\nabla_i\nabla_{\j}-\nabla_{\j}\nabla_i)s^{i\j k},\xi_k\rangle=-\langle s^{i\j l}{R_{i\j l}}^k,\xi_k\rangle\;\;\;\;\mbox{for any}\;\;\;\mathbold{s}\in\cc{\O,T\otimes T\otimes T},
\]
where $\langle,\rangle$ is the $L^2$ scalar product. Using \eqref{killing_weak} for replacing the second term on the left hand side, we get
\[
\langle\nabla_i\nabla_{\j}(s^{i\j k}+s^{jki}),\xi_k\rangle=-\langle s^{i\j k}{R_{i\j k}}^l,\xi_l\rangle+\langle\nabla_is^{i\j k},t_{jk}\rangle.
\]
Now, we can write down this equation for other two tensors, $\mathbold{s}^\prime$ and $\mathbold{s}^{\prime\prime}$, whose components are permutations of those of $\mathbold{s}$, namely, $s_{i\j k}^\prime=s_{jki}$ and $s_{i\j k}^{\prime\prime}=s_{kij}$. Adding the $\mathbold{s}^\prime$-equation to the $\mathbold{s}$-equation, and subtracting the $\mathbold{s}^{\prime\prime}$-equation, we get
\eq{
\label{higher_weak_der}
\langle\nabla_i\nabla_{\j}s^{i\j k},\xi_k\rangle=\langle s^{i\j k},{R_{ki\j}}^l\xi_l-\nabla_it_{jk}-\nabla_{\j}t_{ki}+\nabla_kt_{i\j}\rangle,
}
which means that the second weak derivative $\vec{\nabla}\vec{\nabla}\vec{\xi}\in L^2(\Omega,T\otimes T\otimes T)$ exists:
\[
%\label{higher_weak_der2}
\nabla_{\j}\nabla_i\xi_k={R_{ki\j}}^l\xi_l-\nabla_it_{jk}-\nabla_{\j}t_{ki}+\nabla_kt_{i\j},
\] 
so $\vec{\xi}\in H^2(\Omega,T)$. Note that the product of a weakly differentiable and a smooth function is weakly differentiable. Let us replace $s^{i\j k}$ by $\nabla_ls^{i\j kl}$ in \eqref{higher_weak_der}. We can recast the derivatives of $s^{i\j kl}$ on the other argument of the scalar product on the right hand side because the lower order weak derivatives of $\vec{\xi}$ have already been proved to exist. Hence we conclude that the third weak derivative also exists, moreover, $\vec{\xi}\in H^3(\Omega,T)$. Moving on to the higher derivatives in this manner, we can see that $\vec{\xi}\in\cap_{\ell\geq 1}H^\ell(\Omega,T)$. In order to conclude smoothness from this, we apply the Sobolev embedding theorem, which states that if $2(\ell-r)>n$, where $n$ is the spatial dimension, then $H^\ell(\Omega,T)\hookrightarrow C^r(\cO,T)$, and this embedding is compact, in particular, continuous. 

What remained is the proof that not only $\vec{\xi}\in C^\infty(\cO,T)$, but also $\vec{\xi}\in\cc{\O,T}$. At this point we could refer to the nonintegrability of the Killing equation for a generic metric, but Korn's inequality will be enough. If $n=1$, then $H^1(\Omega,T)\hookrightarrow C(\cO,T)$, implying that the convergence of $\vec{\xi}_k$ is uniform, so it is also pointwise, therefore $\vec{\xi}|_{\partial\Omega}=0$. For $n>1$ the condition $\vec{\xi}|_{\partial\Omega}=0$ does not follow from the convergence of $\vec{\xi}_k$ in $H^1(\Omega,T)$. But recall that the convergence of $\vec{\xi}_k$ and $\K[\vec{\xi}_k]$ means that $\xi\in\dom\cK$. Recall that $\K^+=\cD$, so we have $\cK\!=\!\K^{++}\!=\!\cD^+\,(=\!\D^+)$. Hence $\pp\mapsto\langle\vec{\xi},\D[\pp]\rangle=\smallint_\Omega\xi_i\nabla_{\j}p^{i\j}=\oint_{\partial\Omega}\xi_{(i}n_{j)}p^{i\j}-\int_\Omega\nabla_{(i}\xi_{j)}p^{i\j}$ is continuous. The vector $\vec{n}$ is the outward normal to $\partial\Omega$. Since $\pp|_{\partial\Omega}$ can be any smooth function so that $\|\pp\|$ is arbitrarily small, we have $\xi_{(i}n_{j)}=0$ on $\partial\Omega$, which implies that $\vec{\xi}|_{\partial\Omega}=0$.

Since $\partial\Omega$ is smooth, there are coordinates in a neighborhood $U$ of $x$ such that $\partial\Omega$ is an $n-1$ dimensional plane. Let $\omega$ be a like a mercury droplet sitting on a horizontal pane of glass. That is, $\omega\subset U\cap\cO$ is a domain with a smooth boundary, and there is a ball $B$ of nonzero radius such that $\partial\omega\cap\partial\Omega\cap B\neq\emptyset$. For smooth vector fields on $\omega$ that vanish on $\partial\omega\cap B$, Korn's inequality holds without the term proportional to $\|\vec{\xi}\|^2$ in \eqref{korn}.\cite{chen_jost} Since $\supp\mathbold{t}\subset\Omega$, if $\omega$ is sufficiently small, $\mathbold{t}|_\omega=0$. Thus $\vec{\nabla}\vec{\xi}|_\omega=0$, which -- together with $\vec{\xi}|_{\partial\Omega}=0$ -- implies that $\xi|_\omega=0$, therefore $\xi\in\cc{\O,T}$.   
\end{proof}

\section{Linear elasticity and the momentum constraint}
\label{elasticity_app}
Let us fix the metric, and say that we want to solve the momentum constraint $\nabla_ip^{i\j}=0$ for $\pp$ in some domain $\Omega$ of the $n$ dimensional space. Is there some simple expression of $\pp$ that can be freely prescribed apart from certain regularity conditions which come from the requirement that $\pp\in H^1(\Omega,T\!\vee\!T)$ or $\pp\in C^\infty(\Omega,T\!\vee\!T)$? The simplest guess, the trace of $\pp$ is actually a promising candidate. The resulting equation is the equilibrium equation in linear elasticity: $\nabla_i\sigma^{i\j}+f^{\j}=0$. The traceless part of $\pp$ plays the role of the unknown stress tensor $\mathbold{\sigma}$, and the force density $\vec{f}$ corresponds to $1/n\,\vec{\nabla}p$, where $p=p^i_i$ is the prescribed trace.  
In the Euclidean space the convolution of $\vec{f}$ with Thomson's solution gives the deformation $\vec{u}$ in $\mathbb{R}^n$. The Thomson kernel can be used to prove that there exists a smooth solution $\pp$ if $p$ is smooth. In a curved space we do not have the luxury of having such a fundamental solution at hand. But Korn's inequality, which was initially motivated by linear elasticity, is helpful even in this case. The original inequality underwent several improvements: not only flatness of the metric has been relaxed in \refonline{chen_jost}, but also the strain -- the Lie derivative of the metric along the deformation -- has been replaced by its traceless part in \refonline{dain1}. This is the relevant inequality in this section. 

As usual, it is easier to prove the existence of a traceless solution $\sS[\vec{f}]$ to $\nabla_iS^{i\j}[\vec{f}]+f^j=0$ if the derivative $\nabla_iS^{i\j}[\vec{f}]$ is defined only weakly. Here the symbol $\sS[\vec{f}]$ just emphasizes that we are looking for an assignment of solution to the source $\vec{f}$. This assignment preferably has some continuity property. The regularity on $\vec{f}$ can also be lowered if the differential operator is only weakly defined, and ultimately we are seeking an everywhere defined continuous solution assignment $\sS:L^2(\Omega,T)\to L^2(\Omega,T\!\vee\!T)$. If there are nontrivial (weak) solutions $\tilde{\pp}$ to the homogeneous equation $\nabla_i\tilde{p}^{i\j}=0$, then this assignment is not unique. The existence of such nontrivial solutions in a bounded domain is essential in linear elasticity. This is the freedom that makes various boundary conditions imposed on the stress and the deformation admissible. The most natural way to resolve this ambiguity is to project out the solutions to the homogeneous equation. We demand that $\sS[\vec{f}]$ be orthogonal to all the weak solutions to $\nabla_i\tilde{p}^{i\j}=0$. 

This choice is natural in linear elasticity as well since it corresponds to a simple boundary condition on the deformation $\vec{u}$, as we will see soon. However, the elastic material whose equilibrium equations are analogous to the momentum constraint is rather exotic. As noted in \refonline{dain1}, the corresponding equations of linear elasticity can be derived by varying the energy $E$ with respect to $\vec{u}$, where the energy is defined by $4E=\|\tilde{\mathcal{L}}_{\vec{u}}\g\|^2-4\langle\vec{f},\vec{u}\rangle$ in which $\tilde{\mathcal{L}}_{\vec{u}}\g\define\mathcal{L}_{\vec{u}}\g-2/n\,\g\nabla_iu^i$ is the conformal Killing operator. In the Euclidean space this energy is invariant not only under isometries, but also similarities. That indicates that the bulk modulus of the material is zero. As far as we know, there is no elastic material that is infinitely compressible, but resists shearing. Nevertheless, we can imagine that this fictitious material is glued to the boundary $\partial\Omega$ of the bounded domain $\Omega$. Then the equilibrium equation reads as $\nabla_i(\tilde{\mathcal{L}}_{\vec{u}}\g)^{i\j}+f^j=0$ with the boundary condition $\vec{u}|_{\partial\Omega}=0$. Thus $\tilde{\pp}=\tilde{\mathcal{L}}_{\vec{u}}\g$ is in the image of the conformal Killing operator defined on vector fields vanishing on the boundary, hence it is orthogonal to the kernel of the adjoint of this operator, which is nothing but the weakly defined differential operator $\nabla_i\tilde{p}^{ij}$ with no boundary condition on $\tilde{\pp}$. So if we accept the idea of our exotic elastic material, we have no physical ground for doubts as to the existence of the map $\sS$ with the specification that $\ran\sS$ is orthogonal to the (weak) solutions to the homogeneous equation. The upcoming lemma gives a proof for this expectation. 

Finally, we note that it is advantageous to introduce the deformation $\vec{u}$ even in the investigation of the momentum constraint, where it is only an auxiliary variable. The resulting second order equation $\nabla_i(\tilde{\mathcal{L}}_{\vec{u}}\g)^{i\j}+f^j=0$ for $\vec{u}$ is elliptic, so the arsenal developed for elliptic equations can be deployed.\cite{choquet} Alternatively, one can use a generalization of the Lax-Milgram theorem after proving that its conditions hold, for which Korn's inequality can be used.\cite{dain2} We will not follow these paths here, since we do not need so strong results as the ones that can be obtained by these techniques. Our existence theorem is only local, but it suffices for our goals, and the only tool its proof needs is Korn's inequality. 
\begin{lem}
\label{supmom_sol_lem}
Let the space be at least three dimensional, $\Omega$ and $\D$ as in the previous lemma. Define $\E=\D|_{\tilde{L}^2(\Omega,T\vee T)}$, where $\tilde{L}^2(\Omega,T\!\vee\!T)\subset L^2(\Omega,T\!\vee\!T)$ is the space of traceless tensors. There exists a linear map $\sS:L^2(\Omega,T)\to\tilde{L}^2(\Omega,T\!\vee\!T)$ such that $\dom\sS=L^2(\Omega,T)$, $\ran\sS\subset\dom\cE\cap(\ker\cE)^\perp$, and $\cE\circ\sS=\mathrm{id}_{L^2(\Omega,T)}$. This map is unique and continuous. Furthermore, $\ran\sS^+\subset\sob{1}(\Omega,T)$.
\label{supmom_solution_lem}
\end{lem} 
\begin{proof}
If $\sS$ exists, it is unique. The Killing operator $\K$ was introduced at the beginning of the proof of the previous lemma. Here we are going to use the conformal Killing operator: $\K\!:\!L^2(\Omega,T)\!\rightarrowtail\!\tilde{L}^2(\Omega,T\!\vee\!T)$, $\mathop{\mathrm{dom}}\K=\cc{\Omega,T}$, $\K[\vec{\xi}^{}]\!=\!-1/2\,\tilde{\mathcal{L}}_{\vec{\xi}}\g$, where $\tilde{\mathcal{L}}_{\vec{\xi}}\g\define\mathcal{L}_{\vec{\xi}}\g-2/n\,\nabla_{\!i\,}\xi^i\,\g$. Similarly to the former lemma, $\K^+=\cE$. 

The conformal version of Korn's inequality\cite{dain1} states that if the space is at least three dimensional, there is a constant $C$ such that
\[
%\label{korn_conf}
\|\vec{\nabla}\vec{\xi}^{}\|^2\le C(\|\vec{\xi}^{}\|^2+\|\tilde{\mathcal{L}}_{\vec{\xi}}\g\|^2)
\]
for any $\vec{\xi}\in C^\infty(\cO,T)$ (and hence for any $\vec{\xi}\in H^1(\Omega,T)$). As in the proof of the previous lemma, Friedrichs' inequality and this one imply that for sufficiently small $\Omega$ there is a $\gamma$ such that
\eq{
\label{sobolev_conf}
\|\vec{\xi}\|_{H^1(\Omega,T)}^2\define\|\vec{\xi}\|^2+\|\vec{\nabla}\vec{\xi}\|^2\le\gamma\|\tilde{\mathcal{L}}_{\vec{\xi}}\g\|^2,
}
holds for any $\vec{\xi}\in\cc{\O,T}$ (hence for any $\vec{\xi}\in\sob{1}(\Omega,T)$). Thus $\ker\K=\{0\}$. (In fact, $\ker\cK=\{0\}$.) This is true for any metric, including the ones that admit conformal Killing vector fields because of the additional condition that the vector field vanishes on the boundary $\partial\Omega$. Hence $\K$ is invertible on $\ran\K$. Because of \eqref{sobolev_conf}, this inverse is an $L^2$-continuous linear map, so it can be extended to a continuous linear map on $\overline{\ran\K}$. Therefore we have a continuous operator $\mathbold{R}$ on the entire $\tilde{L}^2(\Omega,T\!\vee\!T)$ which is defined on the closed subspace $\overline{\ran\K}$ by this extension, and on the orthogonal complement $\ker\mathbold{\bar{E}}=\ran\K^\perp$ by zero:
\[
\mathbold{R}:\tilde{L}^2(\Omega,T\!\vee\!T)\to\sob{1}(\Omega,T)\subset L^2(\Omega,T),\;\;\mathbold{R}\circ\K=\mathrm{id}_{\scc{\O,T}},\;\;\ker\mathbold{R}\supset\ker\mathbold{\bar{E}},\;\;\mathbold{R}\;\mbox{is continuous.}
\]
We indicated that $\ran\mathbold{R}\subset\sob{1}(\Omega)$. This follows from \eqref{sobolev_conf}, which also implies that $\mathbold{R}$ is continuous even as map into $\sob{1}(\Omega)$, but we need only its $L^2$-continuity. 

Take the adjoint of $\mathbold{R}\circ\K$. All the adjoints are taken here with respect to the $L^2$ scalar product. If $A$ and $B$ are two densely defined operators so that $AB$ is also densely defined, then usually only $(AB)^+\supset B^+A^+$ holds, but if $A$ is continuous and everywhere defined, then $(AB)^+=B^+A^+$. This is our case, so $(\mathbold{R}\circ\K)^+=\K^+\circ\mathbold{R}^+$. Being the adjoint of a continuous operator, $\mathbold{R}^+$ is also continuous, and $\dom\mathbold{R}^+=L^2(\Omega,T)$. On the other hand, $(\mathbold{R}\circ\K)^+=(\mathrm{id}_{\scc{\O,T}})^+=\mathrm{id}_{L^2(\Omega,T)}$. Since $\K^+=\cE$, we finally have $\cE\circ\mathbold{R}^+=\mathrm{id}_{L^2(\Omega,T)}$. This also means that $\ran\mathbold{R}^+\subset\dom\cE$. As usual, $(\ran\mathbold{R}^+)^\perp=\ker\mathbold{R}^{++}=\ker\mathbold{R}\supset\ker\cE$. ($\mathbold{R}^{++}=\mathbold{R}$ because $\mathbold{R}$ is continuous, everywhere defined.) 
Thus $\ran\mathbold{R}^+\subset\overline{\ran\mathbold{R}^+}\subset(\ker\cE)^\perp$. The map $\sS\define\mathbold{R}^+$ is exactly what we were looking for.
\end{proof}                
\begin{cor}
Let the space be at least three dimensional, $\Omega$ and $\E$ as in the above lemma. Let $L:L^2(\Omega,T\!\vee\!T)\to\mathbb{R}$ be a continuous linear map, and $\ker\cE\subset\ker L$. Then there is an $f\in L^2(\Omega)$ such that $L[\pp]=\langle f,p\rangle$ for any solution $\pp$ to the constraint equation $\mathbold{\bar{E}}[\mathbold{\tilde{p}}]+1/n\,\vec{\nabla}p=0$, where $p\define {p^i}_{\!\!i}\in H^1(\Omega)$ and $\tilde{\pp}\define\pp-1/n\,p\,\g\in\dom\cE$.
\end{cor}
\begin{proof}
Continuity of $L$ implies that $L=\langle\mathbold{t},\cdot\rangle$ with some $\mathbold{t}\in L^2(\Omega,T\!\vee\!T)$. It is enough to prove the statement for traceless $\mathbold{t}$, so we assume that $\mathbold{t}\in\tilde{L}^2(\Omega,T\!\vee\!T)$. By the assumption on $L$, $\mathbold{t}$ is orthogonal to $\ker\cE$. Any solution $\pp$ to the constraint equation can be written as $\pp=\sS[1/n\,\vec{\nabla}p]+\tilde{\mathbold{q}}$, where $p={p^i}_{\!\!i}$, the map $\sS$ is the solution assignment found in the lemma, and $\tilde{\mathbold{q}}\in\ker\cE$. Note that $\sS^+$ is  defined everywhere on $\tilde{L}^2(\Omega,T\!\vee\!T)$. We write $L[\pp]=\langle\mathbold{t},\sS[1/n\,\vec{\nabla}p]+\tilde{\mathbold{q}}\rangle=\langle\mathbold{t},\sS[1/n\,\vec{\nabla}p]\rangle=\langle\sS^+[\mathbold{t}],1/n\,\vec{\nabla}p\rangle$ for any $\pp$ that satisfies the constraint equation. Let $\vec{v}\define 1/n\,\sS^+[\mathbold{t}]\in\sob{1}(\Omega,T)$. $\sob{1}(\Omega,T)$ is the completion of $\cc{\O,T}$ in $H^1(\Omega,T)$, so there is a sequence $\vec{v}_k\in\cc{\O,T}$ that converges to $\vec{v}$ in $H^1(\Omega,T)$. By the definition of the weak derivative, $\vec{\nabla}$ can be recast on $\vec{v}_k$. Since $\nabla_iv^i_{\!k}$ strongly converges in $L^2(\Omega)$ to the weak divergence of $\vec{v}$, we have $L[\pp]=\langle f,p\rangle$ with $f=-\nabla_iv^i\in L^2(\Omega)$.
\end{proof}

\section{Embeddability of spaces and the Hamiltonian constraint}
\label{embeddability_app}
If the class of transverse traceless smooth tensor fields on a bounded domain $\Omega\subset\Sigma$ is broad enough, then we can argue that a function $f:\Omega\to\mathbb{R}$ vanishes if the orthogonality condition $\smallint_\Omega f\,\tilde{\mathbold{q}}^{i\j}\tilde{\mathbold{q}}_{i\j}=0$ holds for all fields $\tilde{\mathbold{q}}$ in this class. If the metric is flat on $\Omega$, and $\Sigma$ is at least three dimensional, then the argument is short. For simplicity, let $\Omega\subset\mathbb{R}^n$ ($n\geq 3$). Let $k\in\mathbb{R}^n$ be an arbitrary vector, and choose other two unit vectors $m$ and $n$ so that the three vectors are mutually orthogonal to each other. Then $\tilde{\mathbold{q}}(x)\define e^{k\cdot x}(m\otimes m-n\otimes n)$ gives $\tilde{\mathbold{q}}^{i\j}\tilde{\mathbold{q}}_{i\j}=2\,e^{2k\cdot x}$, where $\,\cdot\,$ is the standard scalar product in $\mathbb{R}^n$. The space generated by such functions is dense in $L^2(\Omega)$, so $f=0$. 

That was quick. We have to slow down if $\Sigma$ is curved since there are no constant vector fields analogous to $k$, $m$, and $n$. It is plausible, but we cannot be sure that the conclusion of the previous paragraph was not merely due to the flatness of the space. Nonvanishing curvature is an obstacle to the construction of constant vector fields by parallel transport. Fortunately, the generalization of the above argument to curved spaces is not long. This statement seems to contradict the length of this appendix. The reason for this extended note is twofold: all the relevant theorems are included so that the reader is not referred to the literature for the precise statements, and the question is put into the context of embeddability of spaces so that what would be only a technical detail in our argument reveals a (disputably\cite{anderson}) important feature of Einstein's gravity.

The freedom in the choice of a transverse momentum determines what $n\geq 2$ dimensional spaces can be thought of as spacial slices of an $n+1$ dimensional vacuum spacetime. To see this, first note that if the spacetime metric solves Einstein's equations, then in particular all the gravitational constraints are satisfied. In terms of the extrinsic curvature $\K$, its trace $K$, and the Ricci scalar $R$ of the spatial metric $\g$, the momentum and the Hamiltonian constraints are the equations $\nabla_i(K^{i\j}-Kg^{i\j})=0$ and $K^{i\j}K_{i\j}-K^2-R=0$, respectively, or in terms of the momentum $\pp=\K-K\g$ and its trace $p$, $\nabla_ip^{i\j}=0$ and $p^{i\j}p_{i\j}-\frac{1}{n-1}\,p^2-R=0$. Actually, the solubility of the constraints is also sufficient for the existence of such an embedding since if $(\Sigma,\g)$ is a smooth three dimensional Riemannian manifold with a smooth symmetric tensor field $\K$ on it so that all the constraints are satisfied, then there is a globally hyperbolic Ricci-flat spacetime with a Cauchy surface whose induced metric and extrinsic curvature are $\g$ and $\K$.\cite{choquet_geroch} So the real question is if we have enough freedom to prescribe $p^{i\j}p_{i\j}-\frac{1}{n-1}\,p^2$ so that the Hamiltonian constraint is satisfied by a transverse momentum for a given spatial metric. The Campbell-Magaard theorem asserts that this can be done in some neighborhood of any point of $\Sigma$ if the latter is an analytical Riemannian manifold, so such $n$ dimensional manifolds can always be locally embedded into a Ricci flat $n+1$ dimensional Riemannian or Lorentzian manifold. Of course this theorem does not provide us with a unique embedding, and the way in which the space is embedded into the ambient spacetime can be further specified. One of the most natural questions is that if the space can be embedded so that it is a maximal slice of the surrounding spacetime, meaning that $p=0$ on the embedded space. This question has been raised and answered in the affirmative in \refonline{chervon_dahia_romero}, however, $\Sigma$ itself was required to be of Lorentzian signature. Note that this is a necessary condition, since the Hamiltonian constraint reads as $\tilde{p}^{i\j}\tilde{p}_{i\j}-R=0$, in which the first term is a positive definite expression of the traceless $\tilde{\pp}$ if $(\Sigma,\g)$ is Riemannian, so the constraint cannot be satisfied for a metric whose Ricci scalar is not nonnegative. (The tilde over $\pp$ is just our usual mnemonic for tracelessness.) 

In the arguments showing that the constraints are satisfiable for a given spatial metric, some of the components of the extrinsic curvature are eliminated through the algebraic equations given by the Hamiltonian constraint (and $p=0$) so that the momentum constraint becomes an equation to which the Cauchy-Kowalevski theorem applies. The reason why analiticity is required and only local results are obtained is that the proofs rely on this theorem. We give a brief summary of the argument in \refonline{chervon_dahia_romero} for a Riemannian space $\Sigma$. For later purposes we will not set the trace of the momentum to zero, but instead it is assumed to be prescribed by an analytic function $t$. We also want to apply the analysis to a slightly more general kinetic term $G_{i\j kl}p^{i\j}p^{kl}$, where $G_{i\j kl}=\kappa\,(g_{ik}g_{jl}+g_{il}g_{jk})-\lambda\,g_{i\j}g_{kl}$ with constants $\kappa>0$ and $\lambda$. So the constraint quadratic in $\pp$ has the form $2\kappa p^{i\j}p_{i\j}-\lambda p^2=s$, where $s$ is a prescribed function. In a neighborhood of a point $x\in\Sigma$ we introduce Gaussian normal coordinates corresponding to a the hypersurface on which say the first coordinate vanishes. Note that if the metric is analytic in a local coordinate system at $x$, then there are Gaussian normal coordinates in which the metric is still analytic. In these coordinates the metric components $g_{1\j}$ are zero for $j=2,\dots,n$, and $g_{11}=1$. We set the components $p^{i\j}$ to zero except for $p^{22}$, $p^{33}$, and $p^{1\j}$ if $j=1,\dots,n$. For this arrangement $n\geq3$ is necessary. Together with $p^i_i=p^{11}+g_{22}p^{22}+g_{33}p^{33}=t$ we have $n+2$ constraint equations for the $n+2$ nonzero components of $p^{i\j}$, so we have no more freedom to specify more components of the momentum. After the elimination of $p^{33}$ through $p^i_i=t$, the quadratic constraint gives the following second order equation for $p^{22}$:
\eq{
\label{p22}
\left(\frac{g_{22}}{g_{33}}\det g_{[2,3]}\right)\big(p^{22}\big)^2+\left(\big(p^{11}-t\big)\frac{\det g_{[2,3]}}{g_{33}}\right)p^{22}+(p^{11}-t)p^{11}+\midsum_{i,j=2}^ng_{i\j}p^{1i}p^{1\j}-\frac{s}{4\kappa}-\left(\frac{\lambda}{4\kappa}-\frac{1}{2}\right)t^2=0,
}
where $g_{[2,3]}\define g_{22}g_{33}-g_{23}^2$. Since $(g_{i\j})_{i,j=2,\dots,n}$ is positive definite, $g_{22}>0$, $g_{[2,3]}>0$, and thus $g_{33}>0$. If the discriminant of this equation is positive, then using one branch of solutions for $p^{22}$, we get the desired form for the momentum constraint: 
\eq{
\label{mom_constraint_standard}
\frac{\partial p^{1\j}}{\partial x^1}=\midsum_{i=1}^n\midsum_{k=1}^na^{i\j}_k(z)\frac{\partial p^{1k}}{\partial x^i}+b^j(z),\;\;\;\;\;z=(x^1,\dots,x^n,p^{11},\dots,p^{1n}),\;\;j=1,\dots,n,
}
where $a^{i\j}_k(z)$ and $b^j(z)$ are analytic functions on a domain that will be described soon.

A function $f$ is analytical on an open set $\Omega\subset\mathbb{R}^m$ if and only if for any compact set $K\subset\Omega$ there are positive constants $M$ and $r$ such that $f\in C_{M,r}(x)$ for any $x\in K$. $f\in C_{M,r}(x)$ means that $f$ is smooth in a neighborhood of $x$, and 
\[
%\label{CMr}
|\partial_\mathbf{j}f(x)|\leq M\;\mathbf{j}!\;r^{-|\mathbf{j}|}\;\;\;\;\;\mbox{for all}\;\mathbf{j}\in\mathbb{N}^m,
\]
where $\mathbf{j}$ is a multiindex, $\mathbf{j}!\define j_1!\dots j_m!$, and $|\mathbf{j}|\define j_1+\dots+j_m$. Now we are ready to quote the relevant version of the Cauchy-Kowalevski theorem. The theorem is often stated without making it explicit how the size of the neighborhood in which the solution must exist depends on the given data. For later purposes, we included such a specification of the radius of convergence, which of course can be seen from the classic proofs of the theorem.\cite{john}  
\begin{theorname} {\rm(Cauchy-Kowalevski)} Let $a^{i\j}_k$ and $b^j$ be real analytic functions at the origin of $\mathbb{R}^{n+N-1}$. Then the system of differential equations
\[
\frac{\partial u^j}{\partial x^1}=\midsum_{i=2}^n\midsum_{k=1}^Na^{i\j}_k(z)\frac{\partial u^k}{\partial x^i}+b^j(z),\;\;\;\;\;z=(x_2,\!...,x_n,u_1,\!...,u_N),\;\;j=1,\dots,N
\]
with initial conditions $u^j=0$ at $x^1=0$ ($j=1,\dots,N$) has a system of real analytic solutions $u^j(x^1,x^2,\dots x^n)$ in the ball $B_\rho(0)$, where $\rho$ depends on $n$, $N$, and on the class $C_{M,r}(0)$ to which all the coefficients $a^{i\j}_k$ and $b^j$ belong. This is the only solution that is real analytic at the origin. 
\end{theorname} 
Equation \eqref{mom_constraint_standard} has a form slightly different from the expression in this theorem: the coefficients are allowed to depend on $x^1$. However, this is not more general than the system in the theorem since one can always extend the set of unknown variables $(u^1,\dots,u^N)$ by a new function $v$ with a further equation $\partial v/\partial x^1=1$ and initial value $v=0$ at $x^1=0$. Then $v$ plays the role of $x^1$. 

Let $x=(0,x^2,\dots,x^n)$. If $s(x)$ is positive, and $t^2(x)$ is sufficiently small, then the discriminant of equation \eqref{p22} is positive at $x$ if $p^{1\j}(x)=0$. So the constraint equations can be reduced to the form \eqref{mom_constraint_standard}. In order to calculate to what class $C_{M,r}(z)$ the coefficient functions $a^{i\j}_k$ and $b^j$ in \eqref{mom_constraint_standard} belong, we have to evaluate the derivatives of the solution of \eqref{p22} with respect to $(x^1,\dots,x^n,p^{11},\dots,p^{1n})$ at $(0,x^2,\dots,x^n,0,\dots,0)$. The derivatives with respect to $(p^{11},\dots,p^{1n})$ result in negative powers of the discriminant of \eqref{p22}. Combining this observation with the Cauchy-Kowalevski theorem, we arrive at the following characterization of local solutions to the constraints:
\begin{lem}
\label{aux}
Let $(\Sigma,\g)$ be an at least three dimensional Riemannian manifold, $(U,\varphi)$ a coordinate chart in which $\g$ is analytic; $M$, $r$, $a$, and $b$ positive constants. $S$ and $T$ are two families of functions on $U$ such that $S,T\subset\cap_{y\in U}C_{M,r}(y)$ in the given local coordinates; and $a<s$, $|t|<b$ for all $s\in S$, $t\in T$. If $b$ is sufficiently small, then there is a neighborhood $N$ of any point of $U$ such that for all $s\in S$ and $t\in T$ there is a smooth symmetric tensor field $\pp$ which satisfies $\nabla_ip^{i\j}=0$, $p^i_i=t$, and $\;p^{i\j}p_{i\j}-\sigma p^2=s\;$ everywhere on $N$, where $\sigma$ is a constant.               
\end{lem}         
This lemma guarantees a class of local solutions to our constraints on the momentum. The following lemma is about the consequences of orthogonality conditions similar to the one mentioned at the beginning of this appendix. 
\begin{lem}
\label{dense_kinetic}
Let $(\Sigma,\g)$ be an $n\geq 3$ dimensional Riemannian manifold, $x\in\Sigma$, $(\Omega,\varphi)$ a coordinate chart around $x$ in which $\g$ and the positive function $s:\Omega\to\mathbb{R}^+$ in part b) are analytic. For any $\varrho>0$ there is a bounded domain $N\subset\Omega$ containing $x$ with the following property. Let $f\in L^2(N)$. Define $F:\mathbb{R}^n\to\mathbb{R}$ by $F|_{\varphi(N)}\!\define\!\sqrt{g}f\circ\varphi^{-1}$ and $F|_{\mathbb{R}^n\setminus\varphi(N)}=0$. Let $\hat{F}$ be the Fourier transform of $F$. 
\begin{itemize}
\item[a)] If $\smallint_Nf\,\tilde{p}^{i\j}\tilde{p}_{i\j}=0$ for any smooth transverse traceless symmetric tensor field $\tilde{\mathbold{p}}$ on $\bar{N}$, then $\supp\hat{F}\cap B_\varrho(0)=\emptyset$, where $B_\varrho(0)$ is the ball of radius $\varrho$ centered at the origin. 
\item[b)] If $\smallint_Nfp=0$ for any momentum $\pp$ that is transverse on $N$ and satisfies $p^{i\j}p_{i\j}-\sigma p^2=s$, where $\sigma$ is a constant, then $\supp\hat{F}\cap B_\varrho(0)=\emptyset$.
\end{itemize}
\end{lem}
\begin{proof}
Let $g:\mathbb{R}^n\to\mathbb{R}$ be the inverse Fourier transform of the characteristic function of $B_\varrho(0)$, which is the ball of radius $\varrho$ centered at the origin. Such a $g$ is analytic, so for any compact set $K$ there are constants $M$ and $r$ such that $g\in C_{M,r}(y)$ for any $y\in K$. Since $g$ is bounded, there is a constant $c$ such that $c+g$ is positive. Let $g_a$ be the translation of $g$ by $a\in\mathbb{R}^n$, so $g_a(y)=g(y-a)$. 

For any $\delta>0$ there is a neighborhood $U$ of $x$ such that $U$, $T\define\{\,0\,\}$, and the family that is defined by $S\define\{\,d\,\}\cup\{\,g_a+c\,|\;|a|<\delta\,\}$ in the given local coordinates satisfy the conditions of lemma \ref{aux}. Here $d$ can be any positive constant. Thus there is a bounded domain $N\subset U$ which has the property that for any $s\in S$ there is a smooth transverse traceless symmetric tensor field $\tilde{\mathbold{p}}$ that satisfies $\tilde{p}^{i\j}\tilde{p}_{i\j}=s$ on $\bar{N}$. 

Now let $f$ be as in part {\it a)}. $\smallint_{\mathbb{R}^n}\!F\,s=0$ for all $s\in S$, and since $S$ contains a constant, this implies that $F$ is orthogonal to all the translations of $g$ by $|a|<\delta$. Therefore
\[
\midint_{\mathbb{R}^n}\d^n\!k\;\hat{g}(k)^*\hat{F}(k)\,e^{ik\cdot a}=\midint_{|k|<\varrho}\d^n\!k\,\hat{F}(k)\,e^{ik\cdot a}=0\;\;\;\;\;\mbox{for all}\;|a|<\delta,
\]
where $\,\cdot\,$ denotes the standard scalar product in $\mathbb{R}^n$. The function $a\mapsto\smallint_{|k|<\varrho}\d^n\!k\,\hat{F}(k)\,e^{ik\cdot a}$ is analytic on $\mathbb{R}^n$ since it is the Fourier transform of a compactly supported square integrable function. Thus it is identically zero because the above equality shows that it vanishes on an open subset of $\mathbb{R}^n$. That means that $\smallint_{|k|<\varrho}\d^n\!k\,\hat{F}(k)\,e^{ik\cdot a}=0$ for any $a\in\mathbb{R}^n$, hence $\hat{F}(k)=0$ for any $|k|<\varrho$.

Part {\it b)} goes along the same lines. Here $T\define\{\,\epsilon\,g_a\,|\;|a|<\delta\,\}$ ($\delta>0$) and $S=\{\,s\,\}$ satisfy the conditions of lemma \ref{aux} together with some neighborhood $U$ of $x$, provided that $\epsilon>0$ is sufficiently small.                
\end{proof}
\section{Proof of theorem \ref{linC}}
\label{proof_app}
We will use the tensor field $\pp$ instead of the momentum $\ppi=\sqrt{g}\,\pp$. The momentum dependence is indicated explicitly, but the metric dependence is suppressed in our notations. The starting point is
\eq{
\label{C_prop} 
C[\alpha,\beta,\pp]\cd\hh[\pp]+C^i[\alpha,\beta,\pp]\cd\hh_i[\pp]=\midint_\Sigma t_{i\j}[\alpha,\beta]p^{i\j}.
}
Let $\pp_{\rho,\sigma}\define\rho\pp+\sigma\tilde{\mathbold{q}}+\g$ on a bounded domain $\Omega\supset\supp\alpha\cup\supp\beta$, where $\rho$ and $\sigma$ are constants, $\pp$ and the traceless $\tilde{\mathbold{q}}$ are smooth symmetric tensor fields. Note that the inverse metric satisfies the momentum constraint: $\nabla_ig^{i\j}=0$. In order for $\pp_{\rho,\sigma}$ to be a momentum, we can demand for example that on the entire space $\Sigma$ it is compactly supported. Recall that we use the word momentum for only kinematically allowed momentum (see Appendix \ref{assumptions_app}), so any momentum is assumed to be smooth. Since the right hand side of \eqref{C_prop} is linear in $\rho$ and $\sigma$, if $C^i[\alpha,\beta,\pp_{\rho,\sigma}]\cd\hh_i[\pp_{\rho,\sigma}]$, which is
\eq{
\label{cihi}
\rho\midint_\Sigma C^i[\alpha,\beta,\pp_{\rho,\sigma}]\,\nabla_{\j}\tilde{q}^j_i+\sigma\midint_\Sigma C^i[\alpha,\beta,\pp_{\rho,\sigma}]\,\nabla_{\j}p^j_i,
}
vanishes, so does the term in $C[\alpha,\beta,\pp_{\rho,\sigma}]\cd\hh[\pp_{\rho,\sigma}]$ proportional to $\rho\sigma$. The latter term is
\eq{
\label{ch}
(n\lambda-2\kappa)\midint_\Sigma p\,C[\alpha,\beta,\tilde{\mathbold{q}}]\,-2\kappa\midint_\Sigma C[\alpha,\beta,\g]\,\tilde{p}^{i\j}\tilde{q}_{i\j},
}
where $p=p^i_i$ and $\tilde{\pp}=\pp-1/n\,p\,\g$. 

We use the operators $\D$ and $\E$ introduced in lemma \ref{supmom_unique_lem} and \ref{supmom_sol_lem}. They map square integrable tensor fields on a bounded domain $\Omega$ into the space of square integrable vector fields. Any limit or convergence will be meant in these spaces. $\cD$ and $\cE$ are the closures of $\D$ and $\E$. Recall that $\pp\in\ker\cD$ means that there is a sequence of $\pp_k$ which are smooth on $\cO$, $\lim_{k\to\infty}\pp_k=\pp$, and $\lim_{n\to\infty}\D[\pp_k]=0$. Since the kernel of a closed operator is closed, $\overline{\ker\D}\subset\ker\bar{\D^{\phantom{\!\!\!A}}}$, but we did not show that $\overline{\ker\D}=\ker\bar{\D^{\phantom{\!\!\!A}}}$, so we have to be a little careful later. A $\pp\in\ker\cD$ is not necessarily smooth, so $C$ or $C^i$ might not be defined on it. Nevertheless, \eqref{cihi} is zero not only for $\pp$ and $\tilde{\mathbold{q}}$ such that $\pp|_\Omega\in\ker\D$ and $\tilde{\mathbold{q}}|_\Omega\in\ker\E$, but also if $\pp|_\Omega\in\ker\cD$ and $\tilde{\mathbold{q}}|_\Omega\in\ker\E$, or if $\pp|_\Omega\in\ker\D$ and $\tilde{\mathbold{q}}|_\Omega\in\ker\cE$, where the evaluation of \eqref{cihi} and \eqref{ch} on a tensor field $\pp\in\ker\cE\subset\ker\cD$ is meant by taking the limit of the integrals along an approximating sequence of momenta $\pp_k$. On the other hand, it is unpredictable what would happen if we tried $\pp|_\Omega\in\ker\cD$ and $\tilde{\mathbold{q}}|_\Omega\in\ker\cE$. In this case \eqref{cihi} might not converge along momenta tending to $\pp$ and $\tilde{\mathbold{q}}$ even if there is nothing pathological about $C^i$. For instance, if they contain higher derivatives of $\pp$ with smooth coefficient functions, there is already enough room for an unpleasant behavior.    

First let $p=0$ and $\tilde{\pp}=\tilde{\mathbold{q}}$. If $\tilde{\mathbold{q}}$ is a smooth tensor field which is transverse and traceless on $\Omega$, then the first term in \eqref{ch} is absent, and we have $\smallint_\Sigma C[\alpha,\beta,\g]\,\tilde{q}^{i\j}\tilde{q}_{i\j}=0$. Now we show that this implies that $C[\alpha,\beta,\g]=0$. Since $\supp\alpha$ and $\supp\beta$ are compact, by using an appropriate partition of unity, $\alpha$ and $\beta$ can always be written as a (finite) sum of compactly supported smooth functions $\alpha_r$ and $\beta_s$ such that $\supp\alpha_r\cup\supp\beta_s\supset\supp C[\alpha_r,\beta_s,\g]$ is covered by one or two disjoint coordinate domains for any $r$ and $s$, depending on how far the supports of $\alpha_r$ and $\beta_s$ are from each other. So we can assume that $\alpha$ and $\beta$ already have this property. Let $(\Omega,\varphi)$ be one of these covering charts, and $K\define\supp\alpha\cup\supp\beta\cap\Omega$. Assume that $\g$ is analytic in the given local coordinates. By part {\it a)} of lemma \ref{dense_kinetic}, for any $\varrho>0$ there is an open covering of $K$, and thus by the compactness of $K$ a finite subcovering which consists of neighborhoods $N$ so that $\varrho$ and $N$ have the properties as in part {\it a)} of the lemma. By the aid of a partition of unity subordinate to the latter subcovering, we decompose $\alpha$ and $\beta$ into the sum of $\alpha_r$ and $\beta_s$. $\smallint_\Sigma C[\alpha_r,\beta_s,\g]\,\tilde{q}^{i\j}\tilde{q}_{i\j}=0$ for any smooth $\tilde{\mathbold{q}}$ that is traceless and transverse on $\O$ (and zero on $\supp\alpha\cup\supp\beta\setminus\O$). Part {\it a)} of lemma \ref{dense_kinetic} with $f=C[\alpha_r,\beta_s,\g]$, the linearity of the Fourier transformation, and the arbitrariness of $\varrho$ imply that $C[\alpha,\beta,\g]=0$. 

So the second term in \eqref{ch} is zero, and the condition on $C$ is that $\lim_{k\to\infty}\smallint_\Sigma p\,C[\alpha,\beta,\tilde{\mathbold{q}}_k]=0$ if $\pp$ is a momentum transverse on $K$, and $\tilde{\mathbold{q}}_k$ is a sequence of traceless momenta such that $\tilde{\mathbold{q}}_k|_\Omega$ converges and $\lim_{n\to\infty}\E[\tilde{\mathbold{q}}_k|_\Omega]=0$. As in the former paragraph, the chart $(\Omega,\varphi)$ is one of the (at most two) coordinate charts that cover $\supp\alpha\cup\supp\beta$, and $\g$ is assumed to be analytic in $(\Omega,\varphi)$. (On the other chart, if there is any, $\pp$ and $\tilde{\mathbold{q}}_k$ are set to zero.) $\Omega$ is chosen so that it satisfies the conditions of lemma \ref{supmom_unique_lem} and \ref{supmom_sol_lem}. Let $p$ be analytic. By part {\it b)} of lemma \ref{dense_kinetic} (actually, there is no quadratic constraint on $\pp$ here) and a decomposition of $\alpha$ and $\beta$ similar to the one applied in the former paragraph, we conclude that the above limit $\lim_{k\to\infty}\smallint_\Sigma p\,C[\alpha,\beta,\tilde{\mathbold{q}}_k]$ vanishes for all analytic $p$. The contribution of $C^i[\alpha,\beta,\pp]\cdot\hh_i[\pp]$ to the right hand side of \eqref{C_prop} is $\smallint_\Sigma C^i[\alpha,\beta,0]\,\nabla_{\j}p^j_i$, which also goes to zero along $\pp=\tilde{\mathbold{q}}_k$. As we have just proved, so does the contribution of $C[\alpha,\beta,\pp]\cdot\hh[\pp]$, which is $\smallint_\Sigma C[\alpha,\beta,\pp]G[\g]$, if $G[\g]$ is analytic in $(\O,\varphi)$. $L_{\alpha,\beta}[\pp]\define\smallint_\O t_{i\j}[\alpha,\beta]\,p^{i\j}=\smallint_K t_{i\j}[\alpha,\beta]\,p^{i\j}$ is a continuous linear functional on the square integrable fields on $\Omega$. The conclusion of this paragraph is that $\ker\cE\subset\ker L_{\alpha,\beta}$ if $G[\g]$ is analytic in $(\O,\varphi)$.

Now we show that if $\pp\in\ker\cD$ and $p\in H^1(\Omega)$, then $\pp\in\ker L_{\alpha,\beta}$. The coordinate chart $(\Omega,\varphi)$ has the properties as before, in particular, it satisfies the conditions of lemma \ref{supmom_unique_lem} and \ref{supmom_sol_lem}, and as before we define $K\define\supp\alpha\cup\supp\beta\cap\Omega$. 
By the corollary of lemma \ref{supmom_sol_lem}, there is an $f_{\alpha,\beta}\in L^2(\Omega)$ such that $L_{\alpha,\beta}[\pp]=\langle f_{\alpha,\beta},p\rangle$ if $\pp\in\ker\cD$ and $p\in H^1(\Omega)$. According to lemma \ref{supmom_sol_lem}, there is a $\pp\in\ker\cD$ for any $p\in H^1(\Omega)$. Since $H^1(\Omega)$ is dense in $L^2(\Omega)$, $f_{\alpha,\beta}$ is unique. Note that $(\alpha,\beta)\mapsto f_{\alpha,\beta}$ is bilinear. What we showed in the former paragraph implies that $f_{\alpha,\beta}=0$ almost everywhere on $\Omega\setminus K$ since the domain $\Omega$ in the previous paragraph could be replaced by any smaller domain containing $K$. Assume that $G[\g]$ is analytic in $(\O,\varphi)$, and if necessary, let us add a constant $c$ to it so that it is negative on $\Omega$. Here we assume that $\kappa>0$. ($\kappa<0$ is the same, but then $G[\g]+c$ should be positive.) If the momentum $\pp$ is transverse on $\Omega$, and $2\kappa p^{i\j}p_{i\j}-\lambda p^2=-\frac{1}{\sqrt{g}}G[\g]$, then all the constraints $\hh_\mu[\pp]=0$ are satisfied on $\Omega$, so $\langle f_{\alpha,\beta},p\rangle=0$. The next step is familiar. By part {\it b)} of lemma \ref{dense_kinetic} and an appropriate decomposition of $\alpha$ and $\beta$, we conclude that $f_{\alpha,\beta}=0$. Since there is some freedom in the constant added to $G[\g]$ in order to make it negative, and $f_{\alpha,\beta}$ depends linearly on this constant, we reached the conclusion $f_{\alpha,\beta}=0$ with the original analytic $G[\g]$. 

In order to clear the way for lemma \ref{supmom_unique_lem}, we have to argue that $\ker\cD\subset\ker L_{\alpha,\beta}$. This statement is stronger than what we proved in the former paragraph since it is imaginable that an element of $\ker\cD$ can be approximated by only such sequences $\pp_k$ for which the derivative $\nabla^{\j}p_k$ of the trace and the divergence $\nabla_i\tilde{p}_k^{i\j}$ of the traceless part are divergent, but the sum $\nabla_i\tilde{p}_k^{i\j}+1/n\,\nabla^jp_k$ tends to zero. For any $\pp\in\ker\cD$ there is a convergent sequence of momenta $\pp_k$ for which $\lim_{k\to\infty}\D[\pp_k]=0$. Write the traceless part $\tilde{\pp}_k$ of $\pp_k$ as the sum of a term in $\ker\cE$ and a term orthogonal to $\ker\cE$. Since $\ker\cE\subset\ker L_{\alpha,\beta}$, if $\tilde{\pp}_k$ is replaced by the latter term, $L_{\alpha,\beta}[\pp_k]$ remains the same as with the original $\pp_k$. So let us do this replacement. Let $\tilde{\mathbold{r}}_k$ be a traceless tensor that solves $\cE[\tilde{\mathbold{r}}_k]+1/n\,\vec{\nabla}p_k=0$, where $p_k$ is the trace of $\pp_k$. Lemma \ref{supmom_sol_lem} guarantees that $\tilde{\mathbold{r}}_k$ exists. Define $\mathbold{r}_k\define\tilde{\mathbold{r}}_k+1/n\,p_k\,\g$. As shown in the previous paragraph, $L_{\alpha,\beta}[\mathbold{r}_k]=0$. By the continuity of the solution assignment in lemma \ref{supmom_sol_lem}, $\lim_{n\to\infty}\mathbold{r}_k=\lim_{k\to\infty}\pp_k$ since $\tilde{\pp}_k$ solves the same equation as $\tilde{\mathbold{r}}_k$ apart from a term which converges to zero as $n\to\infty$. Since $L_{\alpha,\beta}$ is a continuous functional, $L_{\alpha,\beta}[\pp]=\lim_{n\to\infty}L_{\alpha,\beta}[\mathbold{r}_k]=0$. Apply lemma \ref{supmom_unique_lem} to complete the proof, at least for a metric analytic in $(\Omega,\varphi)$.

By properties \eqref{cont_functional_prop},\eqref{anal_prop}, and lemma \ref{analytic_approx_lem}, we can extend the results to any kinematically allowed metric.

\titlelabel{Appendix \thetitle}
\section{}
\label{app_app}
\begin{lem}
\label{lem_supermetric}
If ${G^{kl}}_{\!\!\!\!\!ij\,}$ is an ultralocal concomitant of the metric with symmetry properties ${G^{kl}}_{\!\!\!\!\!ij\,}={G^{kl}}_{\!\!\!\!\!\!j\!\;i\,}={G^{lk}}_{\!\!\!\!\!ij\,}$, then ${G^{kl}}_{\!\!\!\!\!ij\,}=\kappa({\delta^k}_{\!\!i}{\delta^l}_{\!\!\!\j}+{\delta^l}_{\!\!i}{\delta^k}_{\!\!\!\j})-\lambda\,g_{i\!j}g^{kl}$, where $\kappa$ and $\lambda$ are constant.  
\end{lem}
\begin{proof}
Take any $x\mapsto y(x)$ diffeomorphism such that $y(x_0)=x_0$ and $(y_*g_{i\j})(x_0)=g_{i\j}(x_0)$, that is, $D^i_{\j}\define\partial_jy^i(x_0)\in O(n)$. The transformation rule of $\boldsymbol{G}$, whose components are ${G^{kl}}_{\!\!\!\!\!ij\,}=G_{i\j mn}g^{km}g^{ln}$, is
\[
(y_*\boldsymbol{G})_{i'j'}^{k'l'}(x_0)=(D^{-1})^i_{i'}(D^{-1})^j_{j'}D_k^{k'}D_l^{l'}G_{ij}^{kl}(x_0).
\]
Since $\boldsymbol{G}$ is concomitant of the metric (and ultralocal in it), $(y_*\boldsymbol{G}(\g))(x_0)=\boldsymbol{G}(y_*\g(x_0))$. But by our selection of $y$, this implies that $\boldsymbol{G}(x_0)$ is invariant, so from the transformation rule of $\boldsymbol{G}$ we get
\[
D_i^{i'}D_j^{j'}G_{i'j'}^{k'l'}(x_0)=G_{ij}^{kl}(x_0)D_k^{k'}D_l^{l'},
\]     
which means that $\boldsymbol{G}(x_0)$ is an intertwiner between representations of the orthogonal group on symmetric matrices. Since the symmetrized product of two fundamental representations of $O(n)$ decomposes into two inequivalent irreducible representations, Schur's lemma tells us that the space of the intertwiners is only two dimensional, and the formula in the statement already spans a two dimensional space at every point. The coefficients $\kappa$ and $\lambda$ are constant since they are scalars, concomitant of the metric, and the only way that $\xi^i\partial_i\kappa(x_0)=0$ whenever $\nabla_{(i}\xi_{j)}(x_0)=0$ is that $\partial_i\kappa=0$, and similarly, $\partial_i\lambda=0$.
\end{proof}
\begin{lem}
\label{lem_field_dependence}
Let $\G$ be the manifold of the kinematically allowed fields on $\Sigma$, and $G:\G\to C^\infty(\Sigma)$ such that
\eq{
\label{lem_form}
\frac{\delta}{\delta\f}\nn\cd G[\f]=f((\partial_\mathbf{j}\nn)_{|\mathbf{j}|\le J}),
}
where $f$ is a linear function of $\nn\in\cc{\Sigma}$ and its derivatives up to order $J<\infty$ with smooth coefficients. Then $G[\f](x)$ is the function of $\f(x)$ and the derivatives of $\f$ at $x$ up to order $J$.
\end{lem}
\begin{proof}
Let $(\f_\lambda)_{\lambda\in(0,1)}\!\in\!\G$ be a variation. If we multiply the left hand side of \eqref{lem_form} by $\partial_\lambda\f$, and integrate it over $\Sigma$, we get $\partial_\lambda(\nn\cd\,G[\f_\lambda])$. The right hand side of \eqref{lem_form} shows that $\partial_\lambda(\nn\cd\,G[\f_\lambda])=0$ if $\supp\partial_\lambda\f\cap\supp\nn\!=\!\emptyset$ since $f((\partial_\mathbf{j}\nn(x))_{|\mathbf{j}|\le J})=0$ if $x\notin\supp\nn$. It can be shown that by the assumptions made on $\G$ -- see property \eqref{kinematics_prop} in  Appendix \ref{assumptions_app} -- this implies that $\nn\cd\,G[\f]$ depends on $\f$ restricted to any neighborhood that contains $\supp\nn$. Differentiate $\nn\cd\,G[\f_\lambda]$ with respect to $\lambda$, use \eqref{lem_form} for the functional derivative, integrate by parts to get undifferentiated $\nn$ only, then integrate with respect to $\lambda$ as well, exchange the order of the latter integral with the integration over $\S$, and finally note that $\nn$ is an arbitrary $\cc{\Sigma}$ function. The result is 
\eq{
\label{dGdl}
G[\f_1](x)-G[\f_0](x)=\int_0^1\!\!\d\lambda\sum_{|\mathbf{j}|\leq J}C^\mathbf{j}[\f_\lambda](x)\;\partial_\mathbf{j}\partial_\lambda\f_\lambda(x)
}
for any point $x\in\Sigma$. The multiindex $\mathbf{j}$ labels coordinates on $\S$.  

Let $\p\in\G$. By property \eqref{kinematics_prop}, there is a variation $(\f_\lambda)_{\lambda\in(0,1)}\!\in\!\G$ such that $\f_\lambda\!=\!\lambda\p+(1-\lambda)\p_J$ on a neighborhood of $x$, and $\f_1\!=\!\p$ on $\S$. Here $\p_J$ is the Taylor series of $\p$ about $x$, truncated at order $J$. The right hand side of \eqref{dGdl} vanishes for $\f_\lambda$. So $G[\p_J](x)\!=\!G[\f_0](x)\!=\!G[\f_{\lambda=1}](x)\!=\!G[\p](x)$. $G[\f_\lambda](x)$ depends only on $\f_\lambda$ restricted to an arbitrary neighborhood of $x$, and in a neighborhood of $x$ the function $\p_J$ is determined by $\p$ and its derivatives at $x$ up to order $J$.
\end{proof}
\bibliography{mod_grav}
\end{document}